\documentclass[twocolumn,tighten,trackchanges]{aastex63}
\usepackage{mathtools}
\usepackage{txfonts} 
\usepackage{hyperref}
\usepackage{color}
\usepackage{multirow}
\newcommand{\ud}{\mathrm{d}}
\usepackage{gensymb}
\usepackage{comment}

\defcitealias{PaperI}{EHTC~I}
\defcitealias{PaperII}{EHTC~II}
\defcitealias{PaperIII}{EHTC~III}
\defcitealias{PaperIV}{EHTC~IV}
\defcitealias{PaperV}{EHTC~V}
\defcitealias{PaperVI}{EHTC~VI}
\defcitealias{PaperVII}{EHTC~VII}
\defcitealias{PaperVIII}{EHTC~VIII}
\defcitealias{Narayan_2021}{N21}
\defcitealias{Gelles_2021_published}{G21}
\defcitealias{Deglinnocenti_1985}{LD85}

\newcommand{\m}{M87*}
\newcommand{\s}{Sgr\,A*}

\newcommand{\bz}{$\beta_{2,\,n=0}$}%
\newcommand{\bo}{$\beta_{2,\,n=1}$}%

\turnoffedit

\begin{document}

\title{Polarimetry of Black Hole Spin and Axisymmetric Emission Geometry}
\shorttitle{Polarimetric Tracers of Spin and Emission Geometry}

\author[0009-0000-1376-2352]{Vadim Bernshteyn}
\affiliation{Steward Observatory and Department of Astronomy, University of Arizona, 933 N. Cherry Avenue, Tucson, AZ 85721, USA}

\author[0000-0002-7179-3816]{Daniel~C.~M.~Palumbo}
\affil{Center for Astrophysics $\vert$ Harvard \& Smithsonian, 60 Garden Street, Cambridge, MA 02138, USA}
\affiliation{Black Hole Initiative at Harvard University, 20 Garden Street, Cambridge, MA 02138, USA}

\author[0000-0001-9939-5257]{Dominic Chang}
\affil{Center for Astrophysics $\vert$ Harvard \& Smithsonian, 60 Garden Street, Cambridge, MA 02138, USA}
\affiliation{Black Hole Initiative at Harvard University, 20 Garden Street, Cambridge, MA 02138, USA}

\author[0000-0002-7524-5219]{Tintin Nguyen}
\affil{Center for Astrophysics $\vert$ Harvard \& Smithsonian, 60 Garden Street, Cambridge, MA 02138, USA}
\affiliation{Black Hole Initiative at Harvard University, 20 Garden Street, Cambridge, MA 02138, USA}

\begin{abstract}

The Event Horizon Telescope array has measured the average polarimetric spiral morphology of black hole accretion disks using the so-called ``$\beta_2$'' coefficient. This measurement is dominated by the weakly lensed, direct ($n=0$, counting photon half-orbits) image of the disk, providing insight into magnetic field structure around the M87* black hole. 
The future space-based Black Hole Explorer mission is expected to extend these capabilities, allowing for $\beta_2$ measurement of the strongly lensed $n=1$ image. In this paper, we explore the constraints $\beta_2$ measurements would impose on spacetime and plasma geometries using a semi-analytic conical emission model. 
We show that magnetic field orientation with respect to the emission cone qualitatively changes the dependence of $\beta_2$ modes on model parameters. 
For cone-aligned magnetic fields, we find that the leading-order theoretical prediction from equatorial emitters of $\beta_{2,\, n=0} \approx\beta_{2,\, n=1}^*$ is the dominant trend even in off-equatorial models. 
Specifically, we show $\arg(\beta_{2,\, n=0}\,\beta_{2,\, n=1}^*) \approx 4\chi$, where $\chi$ is the angle determining in-cone magnetic field orientation. 
{We also find that $\arg(\beta_{2,\, n=0}\beta_{2,\, n=1})$ minimizes the effect of magnetic field geometry while preserving a dependence on spin that may be used to measure spin direction.}  We find that observations of the photon ring's $\beta_2$ complement the constraint from  direct image polarization, and can greatly constrain  magnetic field structure even with an unknown external rotation measure. 
We also find that black hole spin and emission cone opening angle would not be tightly constrained by $\beta_2$ alone, unless strong priors are placed on the magnetic field geometry.

\end{abstract}

\section{Introduction}

{The Event Horizon Telescope (EHT) collaboration has made horizon-scale images of supermassive black holes commonplace by linking the world with a global sub-mm very-long-baseline interferometry (VLBI) array. Though the first images of Messier 87* (hereafter M87*) were a triumph of engineering and data science, combining interferometric visibilities in the Fourier domain to reconstruct the on-sky image, the downstream theoretical interpretation and black hole parameter estimate remained limited by our understanding of black hole accretion and its relationship to the lensed image \citep{PaperI, PaperII, PaperIII, PaperIV, PaperV, PaperVI}}. 

From the very first treatments of these black hole images, the hunt for measurements of mass and spin has been dominated by morphological signatures in the image, as distant viewers see a ``shadow'' with size and shape specified by these parameters along with the viewing inclination \citep{Bardeen_1973, Luminet_1979, Falcke_2000, Johannsen_2010}. The mass-to-distance ratio of the black hole scales the angular size of lensed features such as the ``inner shadow'' (the direct image of the horizon) and the ``photon ring'' (the sharp ring feature formed from multiply-lensed copies of the accretion disk), whereas spin and inclination warp these features from circles to more asymmetric shapes \citep[see, e.g.][]{Gralla_2019_photonrings, Johnson_2020, Chael_2021}. 
While these features are shaped more by the spacetime than by the astrophysics of the surrounding accretion disk, they are subtle, and not yet well-constrained by existing Event Horizon Telescope images of the near-horizon plasma \citep[see, e.g.][]{PaperIV, SgrA_PaperIV}.

Indeed, EHT theoretical constraints more often {produce} indirect measurements of properties such as the mass, spin, accretion rate, and disk magnetization through comparisons of low-resolution image properties to those extracted from models, such as semi-analytic radiatively inefficient accretion flows (RIAFs) or general relativistic magnetohydrodynamical (GRMHD) simulations \citep[see, e.g.][]{PaperV, PaperVI}. One such image property that is prevalent in images of gently inclined accretion disks is the polarization spiral pattern, expressed using the so-called $\beta_2$ parameter \citep{PWP_2020}. This parameter expresses the overall pitch angle of a spiral pattern in polarization using the phase of a complex coefficient, related directly to the axisymmetric magnetic field through the polarization of synchrotron radiation.

Image polarization, through analyses of $\beta_2$, has been identified with trends in spin in two general forms. {First, studies of GRMHD simulations in \citet{PWP_2020}, and later in semi-analytic treatments of Blandford-Znajek monopolar and parabolic solutions as well as RIAFs, have shown that the large-scale magnetic field structure carries spin information far beyond the ergosphere. 
Regularity conditions at the horizon ultimately emerge as a trend in the electric vector position angle (EVPA) at larger radii that rotates by nearly $90^\circ$ as spin varies from minimal to maximal \citep{BZ_1977, Chael_2023, Hou_2025, Wong_2025_published}. 
These analyses apply for equilibrium, axisymmetric field conditions obeying a family of magnetospheric assumptions which may not hold in the full generality of accretion systems found in nature.
}

{Second, irrespective of the underlying magnetic field, a fixed polarized emitter necessarily encodes spin information as the polarization vector is transported to a distant observer .\edit1{This effect was} first treated for X-ray polarization in \citet{Connors_1980}, and studied for highly lensed near-horizon emission by \citet{Himwich_2020} and \citet{Gelles_2021_published}. 
A previous analysis in \citet{Palumbo_2025} treated spin inference from optically thin synchrotron emission from the Kerr midplane. 
One may treat the problem of spin inference in polarization while allowing for more flexibility in the emission geometry. 
In this paper, we extend the analysis of \citet{Palumbo_2025} to \edit1{emission} off the equatorial plane, and provide a more robust treatment of the broad polarimetric phenomenology of the first two lensed sub-images of the accretion disk.
}

{To do so, we manually set an emitting profile for a source using an analytic model from \cite{Chang_2024} which showed that most GRMHD simulations can be well approximated by a combination of 3 two-dimensional emitting surfaces: the equatorial plane of the accretion disk, and two conic surfaces representing the front and back components of the jet.} 
We target just the lensed polarization properties of one surface, which we modulate in polar angle. {By restricting to a single cone, we maximally isolate the geometric contributions from the different magnitudes of lensing experienced by the first two sub-images traversing distinct polar sweeps on the way to the observer, as well as geometric effects from the location of the emitter and the parameters of the spacetime itself.}

We begin with a review of the cone model and its appearance in polarization in Section \ref{sec:cone}. We then study the dependence of the direct and indirect image $\beta_2$ quantities in a survey of models with different emission and magnetic field geometries in Section \ref{sec:paramsurvey}. We then consider sums and differences of the direct and indirect image $\beta_2$ quantities as a means of removing the contribution from the unknown magnetic field geometry or the unknown rotation measure (respectively) in Section \ref{sec:coords}. We present our conclusions in Section \ref{sec:conclusion}.


\section{Synchrotron Emission from a Cone in Kerr}
\label{sec:cone}
Here we give a brief review of the conical emission models from \cite{Chang_2024}. 
The authors defined multiple models for the time-averaged appearance of M87*, with the full best-fit model consisting of 3 conic surfaces.
The model extends previous semi-analytic models for optically thin emission in the midplane (\citet{Narayan_2021} treated Schwarzschild, while \citet{Gelles_2021_published} and \citet{Palumbo_KerrBAM} extended the model to Kerr). 
This model provides {an} analytic approximation for the time-averaged structure of emission around a black hole, and it matches particularly well with GRMHD models of the M87* black hole.
Since in this paper we focus on a phenomenological study of polarization patterns and symmetries arising from lensing in Kerr, we will only consider models consisting of a single conic surface for simplicity.
We will refer to these models as single-cone emission models hereafter.~\footnote{While the total Stokes $I$ intensity images \edit1{of} the conical models were analyzed in detail in \cite{Chang_2024}, the same models can also be used to produce Stokes $Q$ and $U$ linear polarization images.}

\subsection{Model Setup}

The single-cone emission model comprises both parameters pertaining to the plasma flow and parameters pertaining to the spacetime geometry, and as such is a powerful tool for studying one group of parameters assuming the other group is relatively well-constrained (for a full description of parameters at play see Table \ref{tab:params}). 
In particular, because the emission model parametrizes the plasma relative to the zero angular momentum observer (ZAMO) frame, it is naturally suited to ``all else equal'' studies of the impact of black hole spin. 
In this paper, we mostly focus on inferring spacetime parameters once there is a sufficiently good guess for the magnetic field orientation. Of course, the precise and complete constraining power of the model based on observed data requires one to consider both plasma and spacetime parameters together in one high-dimensional optimization problem. In Section \ref{sec:coords}, we propose two new observables which roughly separate polarization imprints due to spacetime and due to magnetic field configuration, but otherwise we defer performing the full optimization on real data to future studies. {More broadly, the model itself contains no treatment of spatially or temporally varying (turbulent) plasma properties, and the constraints we study from measurements of polarimetric image quantities contain no notion of an instrument model, let alone a full analysis pipeline. Thus,}  we only aim to demonstrate some simple dependencies in the model to gain some intuition about polarization from off-equatorial accretion features, such as the jet base.

\begin{figure*}
    \centering
    \includegraphics[width=1.0\linewidth]{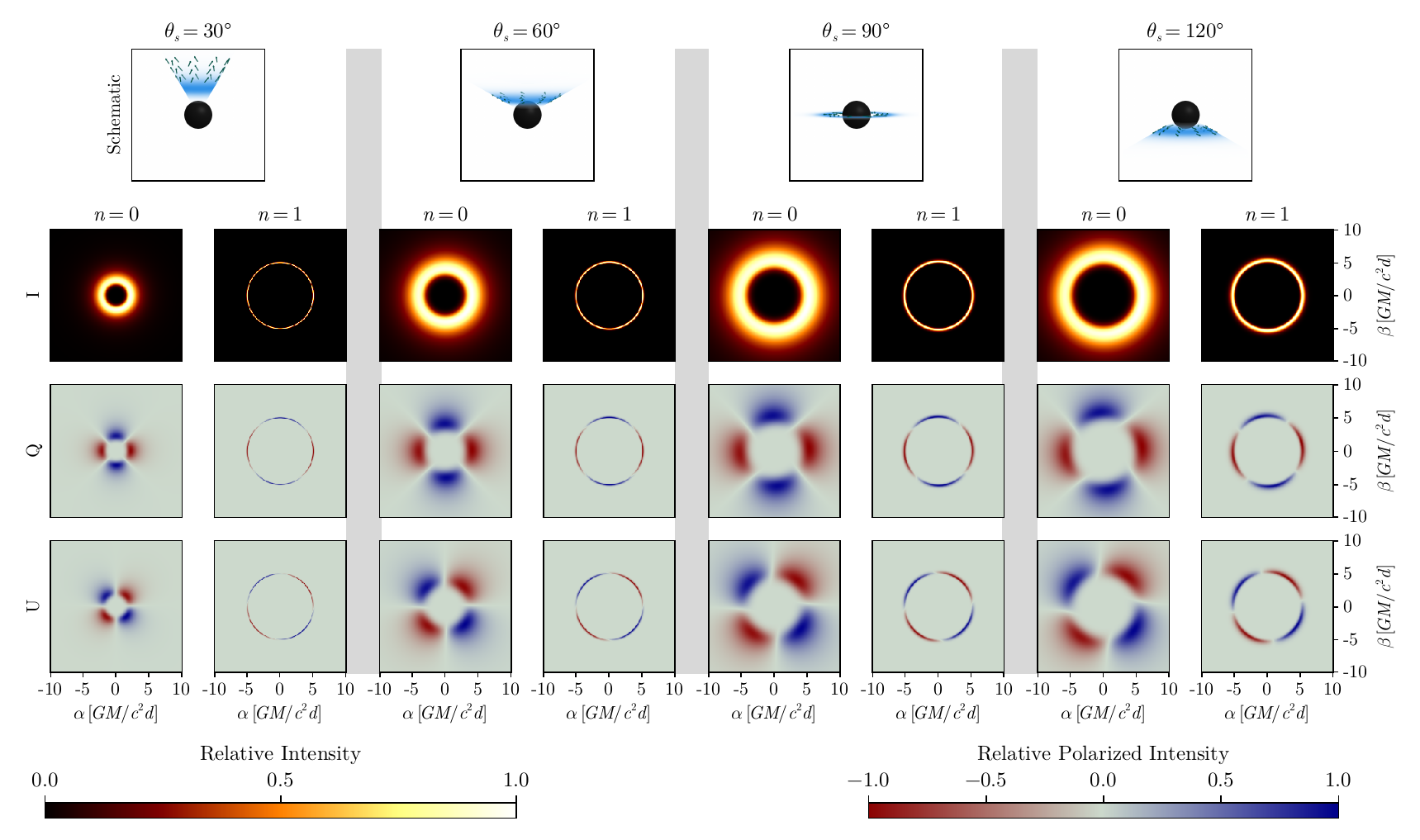}
    \caption{Illustration of single-cone models for different cone opening angles $\theta_s$. Top row shows schematic of emission geometry around a black hole for each $\theta_s$. Rows 2-4 show Stokes I, Q, and U images of $n=0$ and $n=1$ rings for each model, as would be seen by a face-on observer. {Colors represent pixel brightness scaled to the highest value in each individual subplot.} The imaging was done assuming the following {single-cone} model parameters: $a_*=0.9$, $\chi=-90\degree$, $\eta=90\degree$, $\iota=90\degree$, $\beta_v=0.5$, {$R=4$}, $p_1=p_2=4$.}
    \label{fig:model_schematic}
\end{figure*}

We use the \texttt{Krang} software package for numerical ray-tracing of the emission models \citep{Chang_Krang_2024}. 
It is important to mention some limitations of the package. Throughout the paper, a ``face-on'' observer corresponds to {a viewing inclination angle set to $\theta_o=1\degree$}, as the ray-tracing is not numerically reliable  when the angle is set precisely to $\theta_o=0\degree$. Similarly, a ``non-spinning'' black hole corresponds to a numerical model of a black hole with $a_*=0.01$. For our purposes these differences are negligible. 
A schematic of the single-cone model setup and examples of resulting $n=0$ Stokes I images are shown in the top two rows of \autoref{fig:model_schematic}. {The package uses Bardeen $\alpha$ and $\beta$ coordinates on observer's screen, which are dimensionless quantities measured in $\frac{GM}{c^2 d}$, where $M$ is the mass of the black hole and $d$ is the distance to the black hole from the observer.}

Emission \edit1{from} M87* observed by EHT is believed to be produced primarily by synchrotron radiation \citep[see, e.g.][]{Yuan_and_Narayan_2014}. 
As such, it is expected to be linearly polarized perpendicularly to the underlying magnetic field when observed at frequencies at which the emitting material is optically thin, and at a resolution for which beam depolarization is negligible. 
Turbulent sub-structure in the magnetic field can reduce the degree of polarization, a major contributor to differences in polarization morphology among fluid models \citep{PWP_2020}. 
Since the Event Horizon Telescope is expected to greatly improve its polarization sensitivity and resolution with future \edit1{next generation Event Horizon Telescope} \citep[ngEHT,][]{Doeleman_2023} and \edit1{Black Hole Explorer} \citep[BHEX,][]{Johnson_2024} projects, it is interesting to study what insights into the spacetime and emission geometry polarization images provide.

Examples of $Q$ and $U$ signatures of the single-cone models for different cone geometries are shown in the bottom two rows of \autoref{fig:model_schematic}. 
It is clear that the resulting polarization patterns change significantly as the geometry of the emission changes. We also observe that polarization patterns depend on the black hole spin in a manner fundamentally different from the total intensity images, as the orientation of polarization carries additional information about the frame dragging experienced by the emitting plasma. {While we generally defer a full \edit1{3}-surface treatment to future work, we show the analogous breakdown for the complete midplane-and-dual-cone model in \autoref{appendix:dual_cone}.}

The central polarization quantity we investigate in this paper is the $\beta_2$ mode, first introduced in \citet{PWP_2020}. $\beta_2$ is a complex number quantifying rotational symmetry in the observed polarization pattern. Specifically, $|\beta_2|$ quantifies the degree of rotational symmetry in the $P = Q + iU$ image, and $\arg(\beta_2)$ represents the overall pitch angle of the polarization pattern, as each coefficient encodes an integrated projection of the image against simple azimuthal basis functions. In this work, we consider only the fully radially integrated $\beta_2$ coefficient:
\begin{align}
    \beta_2 &=  \dfrac{1}{I_{\rm tot}} \int\limits_0^\infty \int\limits_0^{2 \pi} P(\rho, \varphi) \, e^{- i 2 \varphi} \; \rho \mathop{\ud\varphi}  \mathop{\ud\rho},
\end{align}
where $I_{\rm tot} \equiv \iint_{\rm image} I \, \mathop{\ud x} \mathop{\ud y}$ is the total Stokes $I$ flux density in the image, and $\rho$ and $\varphi$ define radial and azimuthal image coordinates.

$\beta_2$ has been measured from polarimetric observations with the EHT since the first imaging of \m{} and \s{}, and used in downstream theoretical analyses as an image-domain model comparison metric, especially for comparisons to GRMHD \citep{PaperVII, PaperVIII, SgrA_Paper_VII, SgrA_Paper_VIII, EHT2021_2025_published}. 
The coefficient also permits a Fourier analogue described in \citet{Palumbo_2023_b2vis}; as the spiraling polarization handedness is expected to swap between the direct and indirect image \citep{Himwich_2020, Palumbo_2022}, so too does the Fourier analogue flip in sign as the indirect image dominates long baselines. Moreover, the direct and indirect image $\beta_2$ may be independently estimated even with limited Fourier coverage that is otherwise insufficient for imaging \citep{Shavelle_2024, Tamar_2024_published}. The future \edit1{BHEX} space mission \citep{Johnson_2024} is expected to probe the M87* image on scales where the signal from {the} $n=1$ photon ring would dominate the $\beta_2$ signal, thus providing us with two separate $\beta_2$ measurements for $n=0$ and $n=1$. As we demonstrate below, \bz{} and \bo{} show different dependencies on model parameters, meaning measuring both simultaneously would have {greater model-constraining power}.

\subsection{Basic Dependencies of $\beta_2$}
\label{sec:basic}
\begin{figure}
    \centering
    \includegraphics[width=1.0\linewidth]{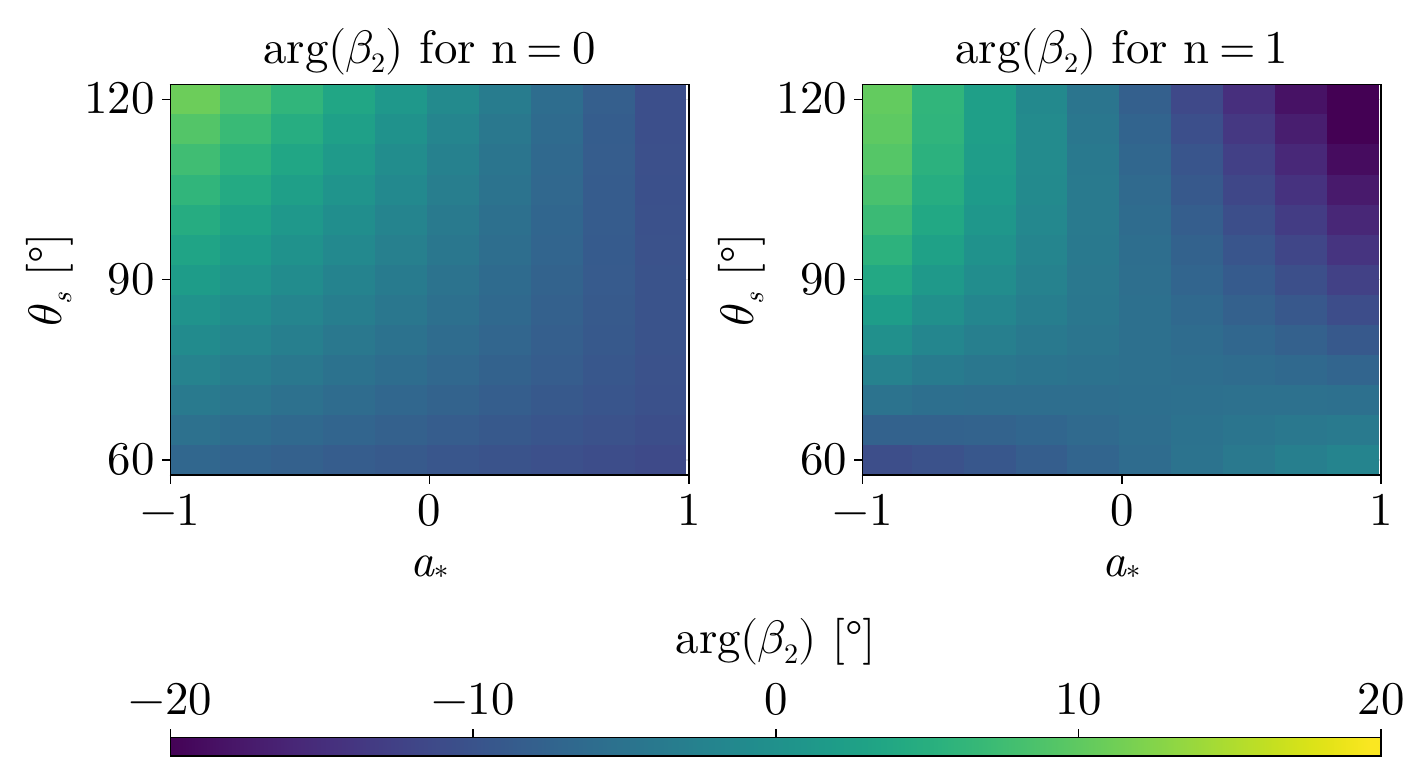}
    \caption{$\arg(\beta_2)$ in degrees as a function of black hole spin $a_*$ and cone opening angle $\theta_s$. Left panel is constructed for $n=0$ rings only, and the right panel is for $n=1$ rings only. The underlying plasma model is: $\chi = -90\degree, \eta = 90\degree, \iota = 90\degree, \beta_v = 0.5, R = 4, p_1 = p_2 = 4$. Imaging was done at observer inclination $\theta_o = 17\degree$.}
    \label{fig:spin_thetas_hm}
\end{figure}

We start by considering \bz{} and \bo{} dependence on dimensionless black hole spin $a_*$ and cone opening angle $\theta_s$ for a single plasma model. We choose the following parameters for our model: $\theta_o = 17\degree, \chi = -90\degree, \eta = 90\degree, \iota = 90\degree, \beta_v = 0.5, R = 4, p_1 = p_2 = 4$. We choose these parameters to approximate the appearance of M87*, for which we assume the spin inclination to match that of the jet observed at large scales \citep{Walker_2018,Kim_2018}. With these parameters fixed, we vary black hole spin $a_*$ and cone opening angle $\theta_s$ in the ranges $a_* \in [-1, 1]$, $\theta_s \in [60\degree, 120\degree]$. Then, for each combination of $a_*$ and $\theta_s$, we rendered $n=0$ and $n=1$ images of the full model using \texttt{Krang}, and computed $\beta_2$ for each image. Finally, we produced heatmaps of $\arg(\beta_2)$ as a function of $a_*$ and $\theta_s$.

The results are shown in \autoref{fig:spin_thetas_hm}. We see that phases of both \bz{} and \bo{} change significantly with $\theta_s$ and $a_*$, which demonstrates that $\beta_2$ modes indeed encode spacetime information. It is especially promising to see that $n=0$ and $n=1$ modes show different patterns in their geometry dependence, such that measuring \bo{} will yield qualitatively new information that is, at least partially, independent of \bz.

\begin{figure*}
    \centering
    \includegraphics[width=0.9\linewidth]{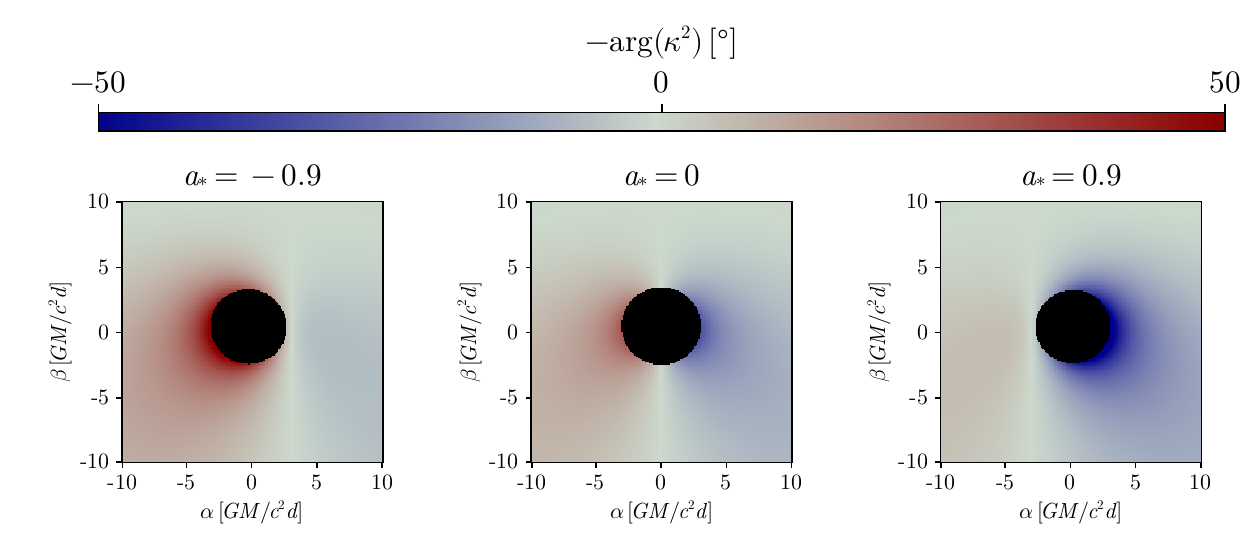}
    \caption{$-\arg({\kappa^2})$ in degrees for each light ray on screen. Images are generated for $n=0$ rings of $\theta_s=90\degree$ cones around black holes with spins {$a_* = -0.9$}, $0$, and $0.9$. All other model parameters are the same as in \autoref{fig:spin_thetas_hm}. The black region satisfies the condition \edit1{that the flux-per-pixel is at least 200 times smaller the maximum flux-per-pixel in the image} {$I_{\text{black}} < \frac{1}{200}\max(I_{\text{image}})$}, and is manually masked out.}
    \label{fig:kappa_img}
\end{figure*}

The precise analytic dependence of the observed $\beta_2$ mode on the black hole spin $a_*$ is sensitive to the geometry of the emitting material, as well as the embedded magnetic field. To better understand the geometric dependencies within the observed polarization, we may turn to the Penrose-Walker constant $\kappa$, a complex number that is conserved along null geodesics in the Kerr spacetime \citep{Walker_Penrose_1970}. Polarization ray-tracing in \texttt{Krang} is done by computing Penrose-Walker constants at each point of emission, and then using these constants to derive polarization of light rays on the observer's screen. The constant mixes dependencies from black hole parameters, the location and magnetization of the emitter, as well as the location of the observer (in setting the direction of the emission of light):
\begin{equation}\label{eq:kappa}
    \kappa = (A - iB)(r - ia_* \cos \theta),
\end{equation}
where
\begin{equation}\label{eq:AB}
    \begin{split}
        A &= k^t f^r - k^r f^t + a_* \sin^2 \theta (k^r f^\phi - k^\phi f^r), \\
        B &= \left[ (r^2 + a_*^2)(k^\phi f^\theta - k^\theta f^\phi) - a_*(k^t f^\theta - k^\theta f^t) \right] \sin \theta,
    \end{split}
\end{equation}
$k^\mu = (k^t, k^r, k^\theta, k^\phi)$ is the null geodesic 4-vector, $f^\mu = (f^t, f^r, f^\theta, f^\phi)$ is the polarization 4-vector, and $(t, r, \theta, \phi)$ are Boyer-Lindquist coordinates at the point of emission. 
While the presence of several distinct orders of spin dependence in $\kappa$ means we ought to expect spin information in downstream polarimetric observations, the particular connection to $\beta_2$ is not obvious by inspection. In an upcoming paper (Nguyen et al., in prep.), it is shown that for modest spins $a_*$ and observer inclinations $\theta_o$, the $\beta_2$ mode can be well-approximated by an intensity-weighted integral over the image-plane value of $\kappa^2$:
\begin{align}
    \label{eq:beta_kappa}
    \beta_2 &\approx \frac{m}{I_{\rm tot}} \int\limits_0^\infty \int\limits_0^{2 \pi} I(\rho, \varphi) \left( \frac{{\kappa}^*(\rho, \varphi)}{|\kappa(\rho,\varphi)|} \right) ^2 \; \rho \mathop{\ud\varphi}  \mathop{\ud\rho}
\end{align}
provided that the linear polarization fraction $m$ is constant across the image; essentially, the argument of $\kappa$ manifests straightforwardly as EVPA, albeit defined with opposite handedness because of the complex conjugation of $\kappa$ in Equation \ref{eq:beta_kappa}. Our single-cone emission models prescribe a narrow emission profile and a constant polarization fraction ($m = 1$), the appropriate regime to use Equation \ref{eq:beta_kappa} to relate $\arg(\kappa^2)$ with $\arg(\beta_2)$. We caution that the latter simplifying assumption is rather unrealistic as more \edit1{physically-motivated} GRMHD simulations have shown that the linear polarization fraction could vary substantially on the image due to turbulence in the accretion flow and optical/Faraday depth effects \citep[see, e.g.][]{PaperVIII, Ricarte_2023}. Nevertheless, our study focuses on the phase of $\beta_2$, which is not affected by the constant polarization fraction prescription. The nature of the connection between $\beta_2$ and black hole spin implicitly expressed through $\kappa$ in Equation \ref{eq:beta_kappa} can be seen in \autoref{fig:kappa_img}. Here we compute $-\arg(\kappa^2)$ for each light ray hitting the screen, assuming we know \edit1{the spin and inclination of each model exactly} (this information is needed for $\kappa$ calculation). We do so for models with $a_*=-0.9$, $0$, and $0.9$. For simplicity, we assume a cone opening angle of $\theta_s = 90\degree$ (equatorial plane), and all other model parameters are the same as in \autoref{fig:spin_thetas_hm}. We see that changing spin tends to rotate $\kappa^2$ for each light ray. While the exact amount of rotation is different for different light rays and different plasma models, the average rotation of $\kappa^2$ on \edit1{the} screen corresponds almost exactly to the rotation of the $\beta_2$ mode. As a result, Equation \ref{eq:beta_kappa} provides insight into how the $\beta_2$ mode, which is purely a property of an image, is connected to the spacetime parameter $a_*$.

\section{Parameter Space Study}
\label{sec:paramsurvey}

For a more detailed exploration of $\beta_2$ trends in the model's parameter space we generate a grid of $124800$ models with different parameter combinations, and for each model we compute $\beta_2$ of the resulting decomposed images. The parameters we used are shown in Table \ref{tab:params}. \bz{} and \bo{} were computed for all specified models by ray-tracing $n=0$ and $n=1$ rings separately.

\begin{table*}
    \centering
    \begin{tabular}{|c|c|c|c|}
        \hline
        Parameter & Values & Number of Values & Description \\
        \hline
         $a_*$ & $-0.91, -0.71, -0.51, \dots, 0.49, 0.69, 0.89$ & $10$ & Black hole spin \\
         $\theta_s$ & $60\degree, 65\degree, 70\degree, \dots, 110\degree, 115\degree, 120\degree$ & $13$ & Cone opening angle\\
         $\theta_o$ & $17\degree$ & $1$ & Observer's inclination angle\\
         $\chi$ & $-180\degree, -165\degree, -150\degree, \dots, 135\degree, 150\degree, 165\degree$ & $24$ & Direction of fluid flow along the cone\\
         $\eta$ & $\chi + 180\degree$ & $1$ & Direction of magnetic field along the cone\\
         $\iota$ & $90\degree, 60\degree, 45\degree, 0\degree$ & $4$ & Direction of magnetic field out of the cone \\
         $\beta_v$ & $0.01, 0.5$ & $2$ & Fluid speed in units of $c$\\
         $R$ & $2, 3, 4, 5, 6$ & $5$ & Characteristic emission radius {in units of $\frac{GM}{c^2}$} \\
         $p_1$ & $10$ & $1$ & Emission profile power \#1\\
         $p_2$ & $10$ & $1$ & Emission profile power \#2\\
         \hline 
    \end{tabular}
    \caption{Parameters used for parameter-space study. For each combination of parameters, a corresponding {single-cone} model was ray-traced. Ray-tracing was done separately for $n=0$ and $n=1$ rings. There are $10 \times 13 \times 1 \times 24 \times 1 \times 4 \times 2 \times 5 \times 1 \times 1 = 124800$ models in the total dataset.}
    \label{tab:params}
\end{table*}

\begin{figure*}
    \centering
    \includegraphics[width=0.9\linewidth]{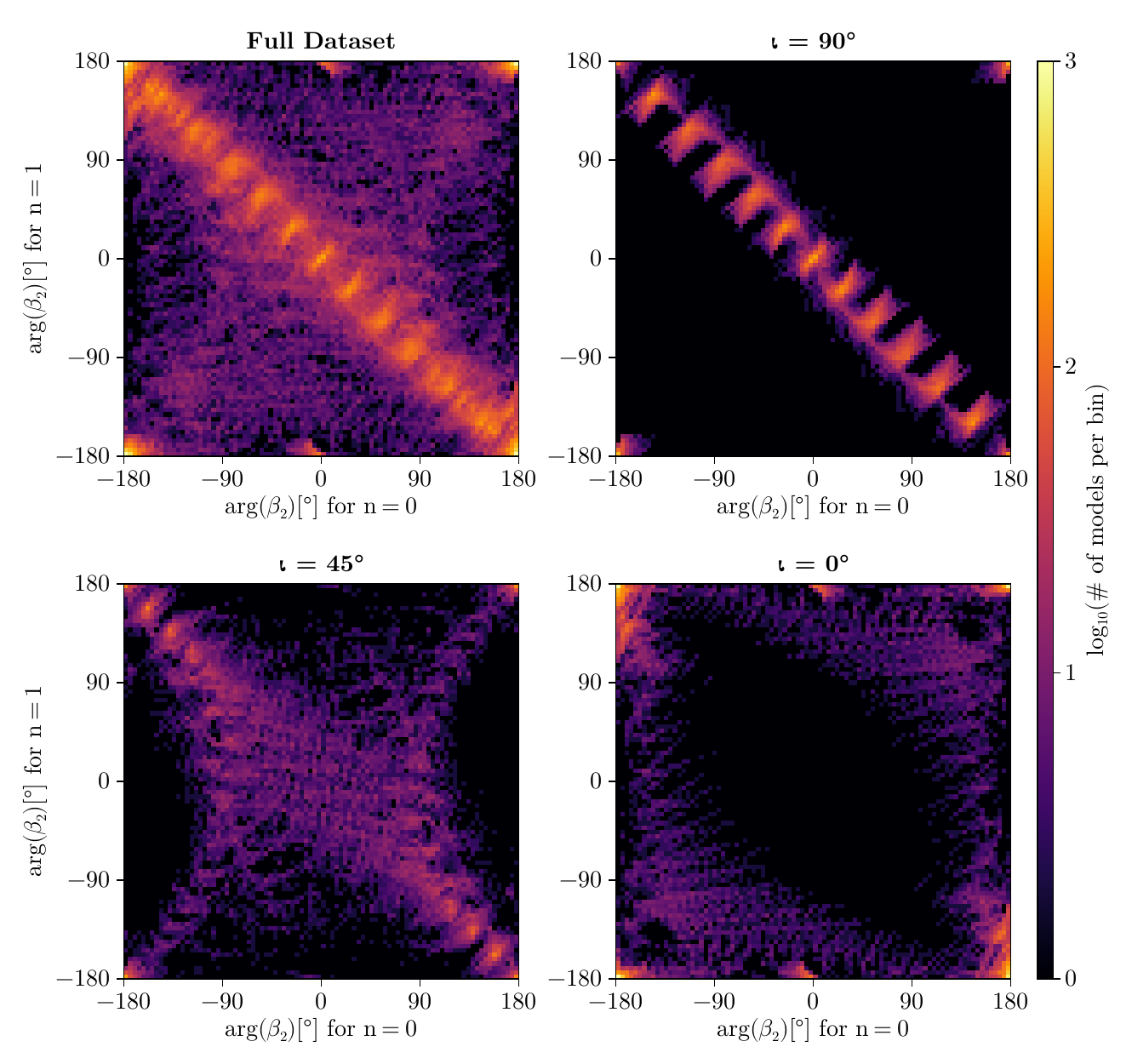}
    \caption{Distribution of \bz{} and \bo{} for considered models. Color corresponds to $\log_{10}$ of the number of models falling into each bin. Size of each bin is $4\degree$ both in \bz{} and in \bo. Note that $\iota = 60\degree$ subset is present in the {``}Full Dataset", but not in other subplots.}
    \label{fig:model_distr}
\end{figure*}

First, note that we do not perform a \edit1{fully} complete sampling of the parameter space, nor is our survey dense enough for inference, meaning the relative densities in $\beta_2$ distributions we obtain should not be taken as precise. That being said, we consider a vast and fairly unconstrained space of models, so we still expect major resulting trends to describe behavior typical for polarized emission in the absence of Faraday effects.

Second, we fix the observer's inclination angle $\theta_o$ at $17\degree$ to mimic the measured inclination of M87*; {though the inferred inclination on jet scales obtained by \citet{Walker_2018} and \citet{Kim_2018} carry order-degree uncertainties, the much larger relevant uncertainty is systematic, in the case where M87*'s spin axis is gently tilted with respect to the jet, a mismatch caused by a possible disk tilt \citep[supported by some evidence of wobble, see, e.g.][]{Cui_2023, Chatterjee_2025}. However, as studied in \citet{Qiu_2023}, the phase of $\beta_2$ is nearly invariant with viewing angle between inclinations of 0 and $\sim30^\circ$, far beyond the reasonable range expected for \m{}; thus, we expect inclination uncertainties to be largely irrelevant to the emergent phenomenology of this work.} 

We restrict $\eta$ to always equal $\chi+180\degree$, as magnetic flux freezing causes magnetic field to point roughly in the direction opposite to the fluid flow by the time the fluid reaches the near-horizon region \citep[see, e.g.][for visualizations of the near-horizon magnetic field structure]{Ricarte_2021}. {While we survey the value of $\iota$, the angle made by the magnetic field with the emission cone, we generally expect $\iota$ to remain close to $90^\circ$, especially for emission surfaces above or below the midplane (that is, corresponding to coronal or jet funnel structures). This expectation follows from general tendencies for plasma motion in these highly magnetized regions, in which particle motion is largely restricted to flow parallel to field lines. These dynamical properties were studied in detail in GRMHD with accompanying field line tracers in both \citet{Mosci_2016} and \citet{Wong_2021}, which together suggest that both in the near-horizon region and at larger distances in the jet, the entrained magnetic field appears mostly parallel to the primary emitting surfaces. We \edit1{also} restrict individual radial contributions:} we \edit1{fix} $p_1=p_2=10$, which corresponds to a very narrow emission profile on the cone, where most of the emission is coming from a circle at distance {$r=R \frac{GM}{c^2}$} from the black hole. We choose narrow emission bands in order to avoid smearing of radially dependent effects within a single emission profile.

In the main body of {this work} we focus on the phase of the $\beta_2$ mode, as this value is expected to be most accurately predicted by the {single-cone} model when compared to real black hole observations, especially those at high frequency with low corresponding Faraday rotation depths. $\arg(\beta_2)$ has also been shown to correlate with the black hole spin in GRMHD simulations \citep{Chael_2023, Emami_2023, Qiu_2023, Wong_2025_published}. By contrast, the amplitude of the mode $|\beta_2|$ is not expected to be predicted accurately by the {single-cone} model due to the assumed (unrealistic) polarization fraction of $m=1$ and the absence of both Faraday rotation and small-scale turbulence, which causes depolarization at finite resolution. 
{Despite this deficiency, comparison} of $|\beta_2|$ between different {single-cone} models might still provide insight into how geometric effects alone affect the observed polarization fraction at finite resolution. 
We find that the angle $\iota$ the magnetic field lines make with the surface of the cone significantly changes the overall polarization of both the direct image and the photon ring. 
We put a more detailed discussion of $|\beta_2|$ trends in Appendix \ref{appendix:beta_mag}.

\subsection{Overall Distribution}
The resulting distribution of $\beta_2$ modes across our models is shown in \autoref{fig:model_distr}. The plot shows $\log_{10}$ of the number of models falling into each bin, where the bins represent specific values of \bz{} and \bo{} measured in the image. The size of each bin is $4\degree$ both in \bz{} and in \bo{}.

We see that most models tend to fall on the main diagonal of the heatmap, which corresponds to $\beta_{2,\, n=0} \approx \beta_{2, \, n=1}^*$. The equality is expected to be exact for pairs of $n$, $n+1$ images for large $n$ from equatorial emission  in Schwarzschild \citep{Himwich_2020}. Models that are strongly offset from the diagonal tend to be produced by exotic orientations of the magnetic field, as can be seen in \edit1{the} $\iota = 45\degree$ and $\iota = 0\degree$ subplots. Overall, low and high values of $\iota$ seem to show qualitatively different patterns, making it difficult to characterize trends in $\beta_2$ without any $\iota$ constraints. Since we expect real black holes to have magnetic field lines approximately tangent to the surface of a jet, we will mostly focus on the $\iota = 90\degree$ case in our analysis. 

We see that $\iota = 90\degree$ models closely follow the $\beta_{2,\, n=0} \approx \beta_{2, \, n=1}^*$ relation. A closer look at the dataset shows that the high clustering of models on a few evenly spaced points on the diagonal is due to sparse sampling in $\chi$, with each cluster corresponding to a specific $\chi$ value imaged. More generally, a detailed analysis of the dataset suggests that $\iota = 90\degree$ models follow two trends in \bz{} - \bo{} space: a location on the main diagonal is primarily determined by the magnetic field orientation $\chi$, and the off-diagonal variation is primarily determined by spacetime parameters $a_*$ and $\theta_s$. This observation suggests a natural change of coordinates, where one coordinate measures the location on the diagonal and the other measures off-diagonal deviation. We explore this new set of coordinates in Section \ref{sec:coords}.

\subsection{Parameter-Specific Trends}
\begin{figure*}
    \centering
    \includegraphics[width=1.0\linewidth]{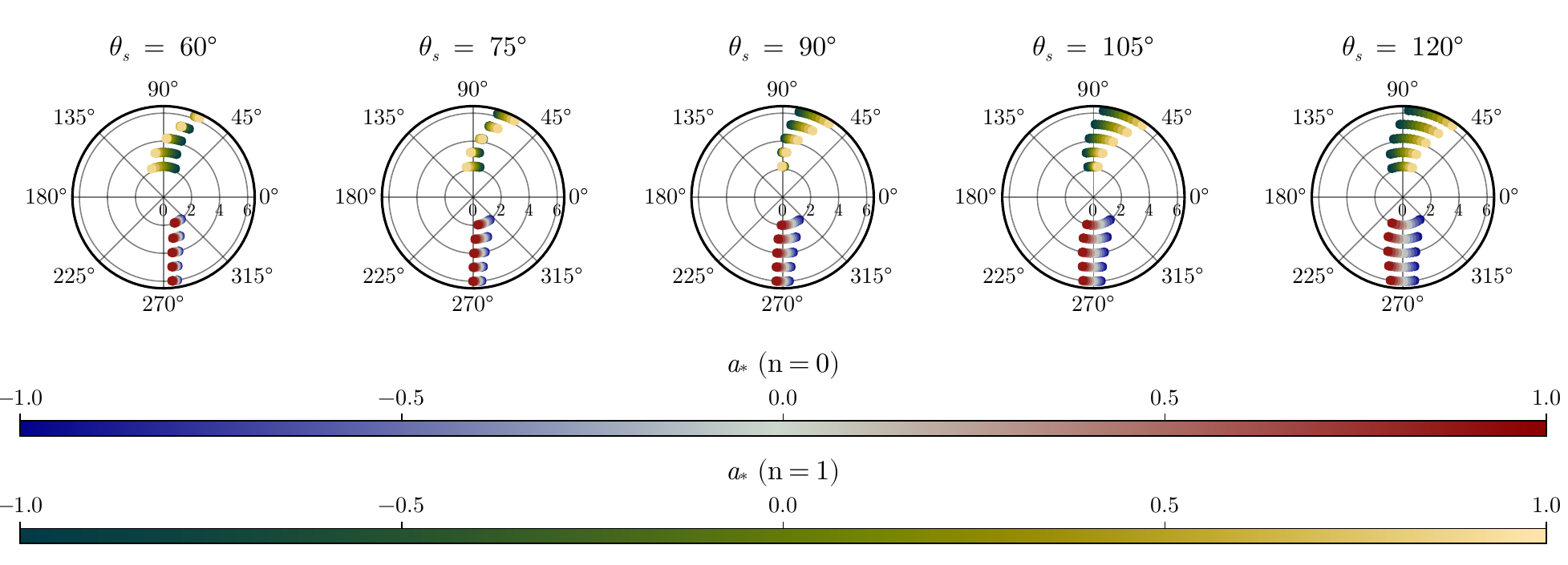}
    \caption{$(R, \, \arg(\beta_2))$ plots in polar coordinates for models with different $\theta_s$ and spin. In each plot, radial coordinate is $R$, and angular coordinate is $\arg(\beta_2)$. All models shown have $\theta_o=17\degree$, $\beta_v=0.5$, $\chi=-135\degree$, $\eta=45\degree$, $\iota=90\degree$. Spin is represented by color, with $n=0$ and $n=1$ measurements using different color scales. The offset in $R$ between $n=0$ and $n=1$ values is set artificially for visual clarity.}
    \label{fig:rotations}
\end{figure*}

Let us now consider specific trends in parameter space in greater detail. First, we fix $\theta_o = 17\degree$, $\chi = -135\degree$, $\eta=45\degree$, {$\iota = 90\degree$}, and $\beta_v = 0.5$. Then we vary $\theta_s$ among the set $[60\degree, 75\degree, 90\degree, 105\degree, 120\degree]$. This way, for each $\theta_s$, the only variables left to be free in our parameter space are emission radius $R$, black hole spin $a_*$, and the ring number $n \in \{0, 1\}$. We depict all of these parameters in \autoref{fig:rotations} in the following way: $R$ is treated as the radial coordinate in a polar-coordinate plot, and $\arg(\beta_2)$ is treated as the phase coordinate. To represent spin, we use color, and we use different color palettes for $n=0$ and $n=1$. 

We see that for $n=0$ rings, larger $\theta_s$ corresponds to larger variance of $\beta_2$ with spin.  This makes intuitive sense, as light from a forward jet undergoes a weaker gravitational lensing effect before reaching the observer compared to a backward jet.
{It also seems that, when averaged over spin, values of \bz{} are largely independent of the value of $R$. This suggests that the observed \bz{} trends would mostly remain the same when wider emission profiles (smaller values of $p_1$ and $p_2$ from Table \ref{tab:params}) are considered.}

For $n=1$ rings, changing $\theta_s$ seems to have almost no {spin-averaged} effect on the dependence of \bo{} on $R$. This suggests that \bo{} might be used to constrain {the} $R$ parameter of the model. Variance of \bo{} with spin also seems to depend on $R$: for small $R$, the variance tends to decrease as $\theta_s$ moves from forward to backward jet, whereas for large $R$ the variance tends to increase.

The trends identified above seem to hold for most models in the parameter space. If this is the case, our results suggest that model fits to data of the {single-cone model} (or spiritually similar approaches)  will constrain spin holistically according to this approximate ``procedure'': First, \bz{} gives an initial idea about the structure of the magnetic field (constraining $\chi$ and $\iota$). Then, the values of $\chi$ and $\iota$ can be used together with \bo{} to constrain $R$, since \bo{} appears nearly independent of $\theta_s$. Knowing $R$ from both total intensity imaging and \bz, we can constrain $\theta_s$ by considering the magnitude of \bz{} rotation at a given $R$. Remaining variation in \bz{} and \bo{} then constrains the spin. In reality, these constraints all occur simultaneously in the Bayesian forward modeling employed by the {single-cone model}.

However, the very first step in this causal chain depends on knowing the phase of $\beta_2$, which is only possible if the external rotation measure is well-measured. Moreover, the knowledge of  \bz{} appears to primarily pin a point along a fundamental diagonal in the space of observed spiral phases, indicating that {parameterizing} spin dependence orthogonal to the diagonal will reduce complexity without losing access to the most interesting physics. Thus, we now turn to phase sums and differences of \bz{} and \bo.

\section{$\Delta$ - $\Sigma$ space}
\label{sec:coords}

\begin{figure*}
    \centering
    \includegraphics[width=1.0\linewidth]{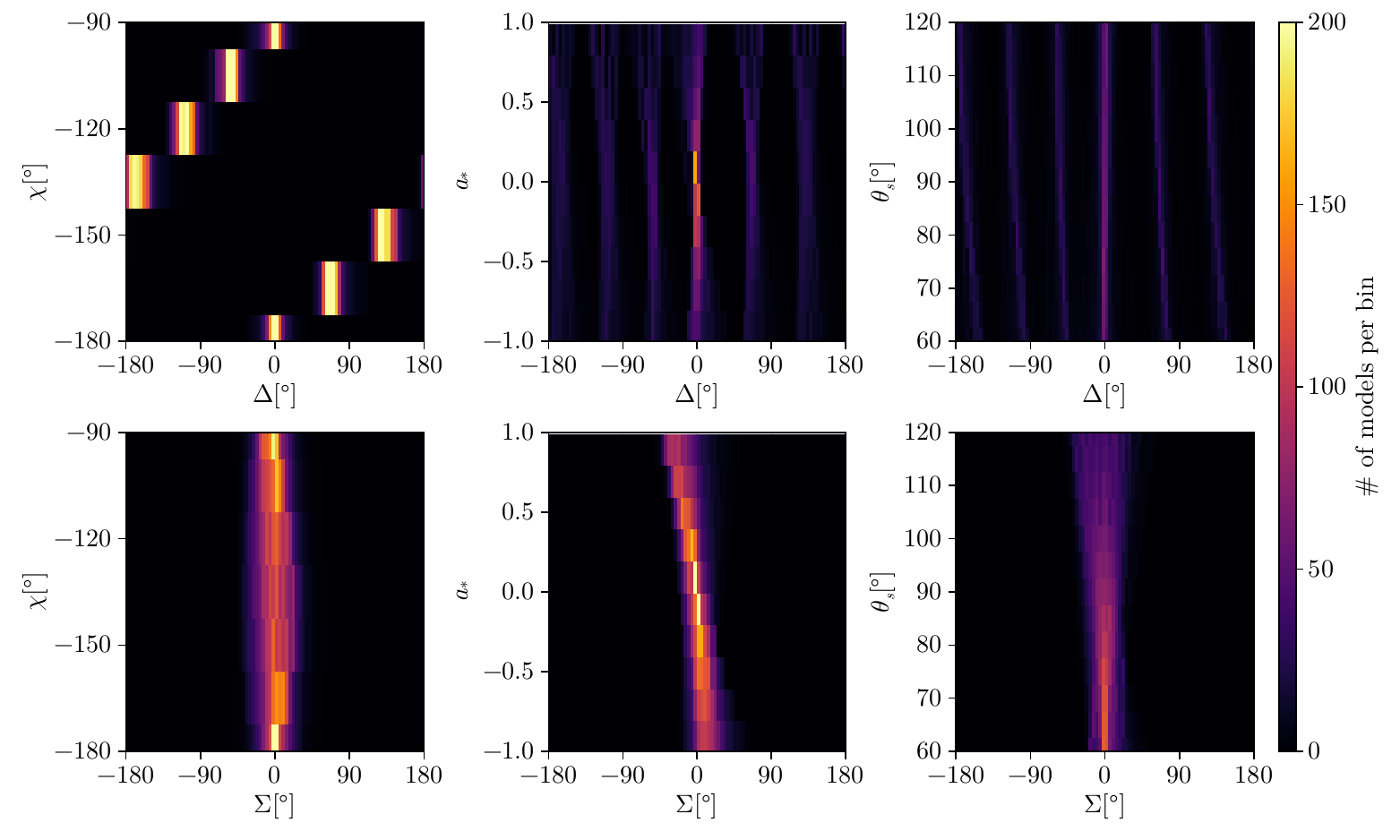}
    \caption{\edit1{Distributions of $\Delta$ and $\Sigma$ in models with $\iota=90^\circ$ and $-180\degree \leqslant \chi \leqslant -90\degree$}. Top: Marginal distribution of $\Delta$ against $\chi$, $a_*$, and $\theta_s$. Bottom: Marginal distribution of $\Sigma$ against the same parameters.}
    \label{fig:chi_constr}
\end{figure*}

We define
\begin{equation}\label{eq:coords}
\begin{split}
    \Delta &\coloneqq \arg(\beta_{2, \, n=0}\beta_{2, \, n=1}^*) = \arg(\beta_{2,\,n=0}) - \arg(\beta_{2,\,n=1}) \\
    \Sigma &\coloneqq \arg(\beta_{2, \, n=0}\beta_{2, \, n=1}) = \arg(\beta_{2,\,n=0}) + \arg(\beta_{2,\,n=1})
\end{split}
\end{equation}
as a basis for \edit1{a} new coordinate system in \bz- \bo{} space. These new coordinates are orthogonal in the apparent parameter space identified in \autoref{fig:model_distr} and as such encode mostly independent information about the model. Specifically, $\Delta$ measures the location along the main diagonal in \autoref{fig:model_distr}, while $\Sigma$ measures the distance from the diagonal ($\Sigma=0$ corresponds exactly to the main diagonal).

Distributions of $\Delta$ and $\Sigma$ with respect to magnetic field orientation, spin, and cone opening angle are shown in \autoref{fig:chi_constr}. We restrict our analysis to models with $\iota=90\degree$, as these are the models where the benefits of the new coordinate system are most apparent. {We examine the deviation from the trends identified in this section later in \autoref{appendix:delta_iota}.} For conciseness and visual clarity, we also assume infalling clockwise-rotating flow ($-180\degree \leqslant \chi \leqslant -90\degree$) {with respect to the spin axis}, even though the trends we discuss are not specific to this range of $\chi$. From the plot we see that $\Delta$ appears to have a strong dependence on the magnetic field orientation, while providing little insight into spacetime parameters $a_*$ and $\theta_s$. On the contrary, $\Sigma$ appears to have a significant correlation with spin, while seemingly having no correlation with $\chi$. This agrees with the intuition that $\Delta$ and $\Sigma$, being orthogonal coordinates quantifying two distinct trends in $\beta_2$ space, should separate magnetic field and spacetime effects as cleanly as possible. We discuss specific trends and observational prospects of each of the coordinates below.

\subsection{$\Delta$ Quantity}
$\Delta$ is a coordinate that follows closely the magnetic field orientation in the model, as can be seen in \autoref{fig:chi_constr} (Top Left panel). In particular, we find that
\begin{equation}\label{eq:chi_delta}
    \Delta \approx 4\chi
\end{equation}
to within $\sim20\degree$ variation in $\Delta$ caused by spacetime effects. It is clear that $\Delta$ is a good predictor of $\chi$ \edit1{(up to $90\degree$ rotation)}, and as such it might be an interesting observable for the BHEX mission.

Future observations of $\Delta$ might be especially useful due to its independence of the external rotation measure. Using traditional \bz{} and \bo{} to constrain \edit1{the} spin of M87* is complicated by the need to assess the magnitude of a rotation measure (RM) in order to accurately measure $\beta_2$ modes \citep[see, e.g.][and its impact on the model scoring in the downstream polarimetric analyses from the 2017 data]{Goddi_2021}. Rotation measure quantifies the extra rotation of polarization pattern induced by the medium along the line of sight. Since $\arg(\beta_{2, \, \text{observed}}) \approx \arg(\beta_{2, \, \text{source}}) + 2\text{RM}\lambda^2$ (with $\lambda$ the observing wavelength, see \citet{Kulsrud_2005}), the accuracy of measuring both \bz{} and \bo{} relies heavily on the accuracy of RM measurement. Previously, it has been shown that in the current EHT setup RM is not tightly constrained \citep[see, e.g.][]{EHT2021_2025_published}, which in turn might weaken constraints obtained from \bz{} and \bo. We note that any remaining difference in the total internal Faraday rotation between $n=0$ and $n=1$ would still be present in $\Delta$, while the external rotation (beyond the near-horizon region) is fully removed by construction. {As studied in  \citet{Jimenez_2021}, the $n=1$ ring can appear depolarized in 230 GHz images of simulations of the M87* accretion flow, even when the $n=0$ image is polarized, largely
\edit1{attributable}
to the increased Faraday rotation depth of the more highly winding rays. However, as described in greater detail in \citet{Palumbo_2022}, in models in which the photon ring is polarized at all, the polarization is generally coherent and exhibits \edit1{a conjugation structure similar to the one found in this study.}}


\subsection{$\Sigma$ Quantity}
$\Sigma$ is, on the other hand, a coordinate that shows correlation with the black hole spin $a_*$, while being relatively independent of the magnetic field orientation (\autoref{fig:chi_constr}, Bottom Left and Bottom Middle). To be precise, the Pearson correlation coefficient between data points shown in the bottom middle panel of \autoref{fig:chi_constr} { is $p \sim -0.6$}, indicating a statistically significant linear dependence between $\Sigma$ and $a_*$. The least-squares fit of this linear dependence gives
\begin{equation}\label{eq:spin_sigma}
    a_* \approx -0.02 (\Sigma / \deg)
\end{equation}
where $\Sigma$ is restricted to $[-50\degree, 50\degree]$ and is measured in degrees. If the relation \edit1{were} exact, \edit1{it} would imply that a $\pm 1\degree$ measurement of $\Sigma$ would constrain the black hole spin $a_*$ to $\pm 0.02$ precision. Put another way, the slope of $\Sigma$ is $\sim50^\circ$ per unit spin. Matching the typical spin imprint magnitude with instrument sensitivity to obtain a spin measurement of $\pm0.1$ precision requires a measurement of the indirect image polarization phase to better than $\sim5^\circ$, corresponding to a signal-to-noise ratio of $\sim 11.5$ in polarimetric visibilities, which is stringent but likely achievable on the most sensitive BHEX baselines \citep{Johnson_2024}. {It bears emphasizing that sensitive polarimetry is generally challenging in VLBI, with stringent needs for high polarimetric purity and parallactic angle coverage in order to successfully calibrate leakage terms \citep[see, e.g.][]{PaperVII}. While the \edit1{orientations} of the $n=0$ and $n=1$ EVPA are generally widely separated in angle for non-radially-infalling flows (and thus more tolerant of small systematic deviations in polarimetric orientation calibration in differential measurements), full degeneracies between gain ratios and linear polarization such as those that emerge in purely linear-feed VLBI could in principle completely destroy the expected signal \citep{TMS}. Thankfully, virtually all proposed sub-mm interferometric expansions currently under consideration either use all-circular feeds (shifting this degeneracy from linear to circular polarization), or have mixed polarimetric feeds.}

Unfortunately, without stronger assumptions on the magnetosphere, the spread of the fit in Equation \ref{eq:spin_sigma} is quite high, as is visually apparent from the gentleness of the slope in the bottom middle panel of \autoref{fig:chi_constr} compared to the apparent spread of the model set. In our dataset the \edit1{standard deviation from} this fit is $\sim 0.47$ units of spin, \edit1{as shown in Appendix \ref{appendix:delta_iota}}. This means that in the absence of additional model constraints a measurement of $\Sigma$ would only roughly constrain the overall direction (into or out of the page) of the \m{} spin vector, which is so far only constrained by brightness asymmetry in GRMHD library comparisons \citep[see, e.g.][]{PaperV}. In addition, unlike $\Delta$, $\Sigma$ is not a relative quantity and as such depends on the external RM measurement. {In the absence of other constraints on the geometry of the emitting material (such as those inferred from image morphology, which is neglected in our analysis), even a highly sensitive polarimetric measurement on BHEX baselines would be dominated by this intrinsic variation among models.}

\section{Conclusions}
\label{sec:conclusion}
In this paper, we studied polarimetric $\beta_2$ modes of {a semi-analytic single-cone model}, focusing on the dependence on cone opening angle, black hole spin, and magnetic field orientation. We did so by sampling the parameter space of the model and ray-tracing corresponding {emission} cones. 
The study was done separately for $n=0$ and $n=1$ rings, as the future BHEX and ngEHT missions are expected to probe M87* on scales where \bz{} and \bo{} modes are separable. We proposed $\Delta =\arg(\beta_{2, \, n=0}\beta_{2, \, n=1}^*)$ as a quantity that is independent of external rotation measure and that can be used effectively to constrain magnetic field orientation $\chi$ in future observations. We also proposed another quantity, $\Sigma =\arg(\beta_{2, \, n=0}\beta_{2, \, n=1})$, as a potential tracer of black hole spin. {Provided that $\iota$ remains within a few tens of degrees of $90\degree$, these two quantities obey simple trends with respect to both magnetic field orientation and spin.} 

In general, we find that the {relationship between} \bz{} and \bo{} and spin is sensitively dependent on the geometric parameters describing the emission location and magnetic field. For cases where magnetic field lines are constrained to the surface of the cone ($\iota = 90\degree$), magnetic field orientation is the primary parameter determining $\beta_2$ for both $n=0$ and $n=1$. The dominant trend is: $\arg(\beta_{2, \, n=0}) = \arg(\beta_{2, \, n=1}^*) = 2\chi - 180\degree$. Deviations from this trend do not exceed $\sim 20\degree$, and are caused by spin and geometric effects. As magnetic field lines move out of the cone's surface, the dependence of $\beta_2$ on magnetic field orientation weakens, and for $\iota = 0\degree$ \bz{} and \bo{} are primarily determined by spin and cone geometry.

Our analysis permits a large volume of potentially unphysical magnetic field geometries \edit1{in} our model survey when constructing spin distributions; this model space stands in contrast to magnetospheric solutions constructed to match expected conditions in a magnetized accretion disk around a realistic black hole \citep[see, e.g.][]{BZ_1977, Chael_2023, Hou_2025, Wong_2025_published}, for which much stronger, order-unity deviations in polarization orientation are expected from black hole spin. More generally, our model makes no connection between the black hole spin and the astrophysical parameters of the model, whereas more realistic disk-jet systems likely carry spin information in the geometry of their emission. In this way, our model may be viewed as pessimistic in its total access to spin information, though it is also optimistic in its simplicity, particularly in its reduction of the emitting region to a single cone.

{It bears emphasizing that the single-cone model studied in this paper leaves out sub-structural details of real accretion disks and their lensed images that will ultimately set a floor in systematic uncertainty of future spacetime measurement efforts. Ideally, models fit to multi-frequency measurements of accretion flows could self-consistently capture the full four-dimensional structure of the plasma as well as the frequency-dependent radiative transfer through the disk and jet in parallel with the spacetime model, resolving turbulence through high-resolution, high-cadence observations. While such models would carry heavy computational expense, they are already under development, and have been applied with heavy assumptions under which the problem is well-posed even with currently available data (such as a midplanar emission geometry with a vertical field, explored in \citet{Levis_tomography} and \citet{Feng_2026}). As such models grow in complexity, their interpretability and pedagogical value decline; thus, we turn to simple models such as the single-cone model treated here to better understand the source of spin information in potential future measurements. Studies that connect these maximally general accretion disk prescriptions to full forward models of an interferometric array \citep[e.g.][]{Tiede_2026} are already underway, and their resulting spin measurement projections will be more decipherable in light of forerunner studies such as ours.}

Ultimately, our analysis provides estimates of the magnitude of a spin imprint in polarization which must be present in time-averaged observations of the first two lensed images at sufficiently high frequency to be optically- and Faraday-thin, while remaining agnostic to many possibly-stronger signatures which depend on astrophysics we leave unspecified. In this way, we provide a fundamental floor on instrumental capabilities for missions targeting these lensed structures.



\appendix

\section{3-Surface Model}\label{appendix:dual_cone}
\begin{figure*}
    \centering
    \includegraphics[width=1.0\linewidth]{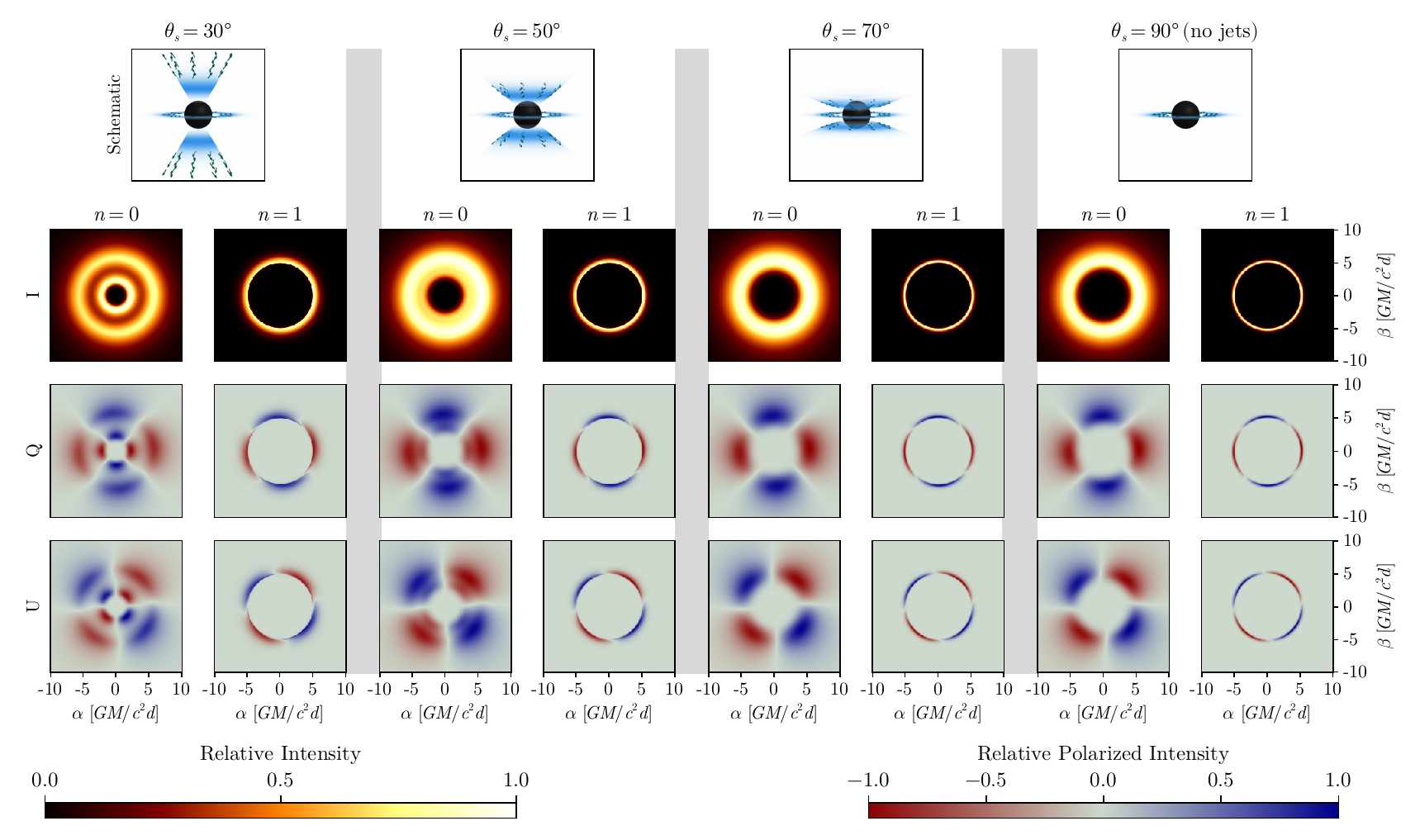}
    \caption{{Same as \autoref{fig:model_schematic}, but for a full 3-surface (dual jet + equator) model. For simplicity, all surfaces share the same plasma parameters as in \autoref{fig:model_schematic}, the top jet has the cone opening angle $\theta_s$, and the bottom jet mirrors it with the opening angle $180^\circ - \theta_s$.}}
    \label{fig:dual_cone}
\end{figure*}

{
The full best-fit model for M87* described in \cite{Chang_2024} consists of 3 emitting conic surfaces (forward jet, backward jet, equatorial plane) which are imaged simultaneously. In this paper we perform a phenomenological study of polarization signatures of a single emitting conic surface, and as such we defer the full 3-surface analysis to a future study. However, since we presume our analysis remains useful even for multi-surface emission provided that optical and Faraday depths in the plasma are negligible, we provide an example of polarized images of the full M87* model.}

{\autoref{fig:dual_cone} shows examples of 3-surface models with varying jet opening angles in a fashion similar to \autoref{fig:model_schematic}. For simplicity, we assume the same plasma parameters for all surfaces we image as those used in \autoref{fig:model_schematic}. We also assume the bottom jet mirrors the top jet, with the cone opening angle of $180^\circ - \theta_s$. We can see that new surfaces add more structure to intensity and polarization images, including a more complicated radial dependence of the signal. However, as \edit1{the} $\beta_2$ coefficient averages the radial structure, it is reasonable to assume based on \autoref{fig:dual_cone} that the major parameter-space trends in $\beta_2$ would at least partially apply to 3-surface models as well, particularly for $\iota$ close to $90^\circ$.}

 \section{$|\beta_2|$ Trends}\label{appendix:beta_mag}
\begin{figure*}
    \centering
    \includegraphics[width=1.0\linewidth]{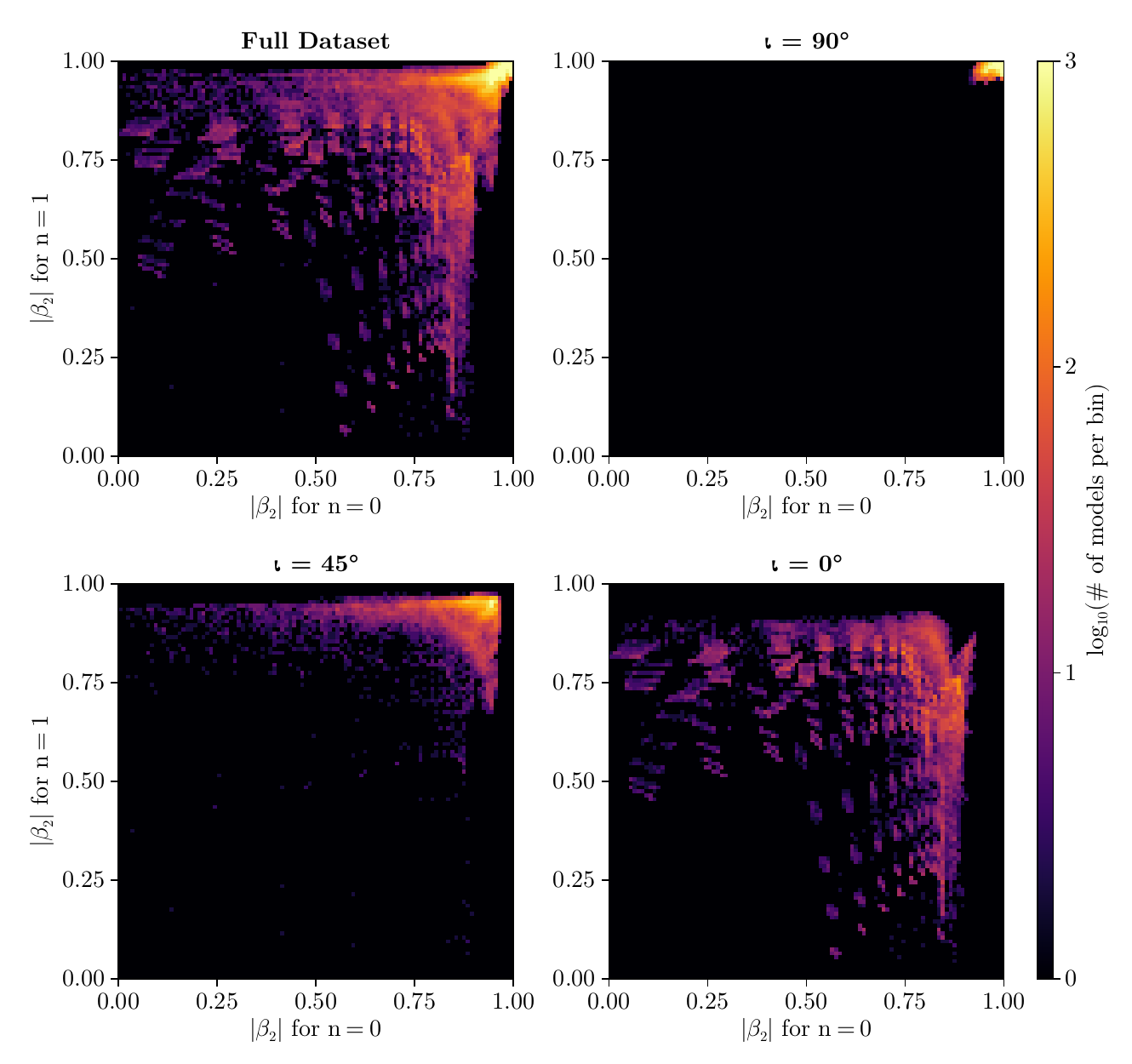}
    \caption{Distribution of models from \autoref{fig:model_distr} in $|\beta_{2, \, n=0}|$ - $|\beta_{2, \, n=1}|$ space. Size of an individual bin is $0.01 \times 0.01$.}
    \label{fig:mag_distr}
\end{figure*}

In the main body of the paper we focused on the phase angle $\arg(\beta_2)$ of the $\beta_2$ modes found in our models. While the magnitude $|\beta_2|$ can be computed using the same numerical procedure, this quantity is expected to carry less physical significance. This is due to the polarization fraction of $m=1$ assumed in {the single-cone} model, which is expected to be unrealistic for plasma emitters around real black holes. As a result, the polarization magnitude predicted in our models is unrealistically high. Moreover, the absence of Faraday rotation effects in \texttt{Krang} ray-tracing calculations increases this disparity even more. However, even though the absolute values of $|\beta_2|$ are physically meaningless, the relative coefficient magnitudes across our surveyed parameter grid can still provide insight into the intrinsic effects of emission geometry on the rotational symmetry of the polarization field. 
As such, we provide $|\beta_2|$ distributions we obtained here. 

Analogous to \autoref{fig:model_distr}, the distribution of all models considered in this study in $|\beta_{2, \, n=0}|$ - $|\beta_{2, \, n=1}|$ space is shown in \autoref{fig:mag_distr}. We see that lower angles of $\iota$ tend to produce weaker $\beta_2$ signatures both in the direct image and in the photon ring. We observe that, even for a totally axisymmetric magnetic field geometry viewed at a small inclination, the degree of rotational symmetry in the observed polarization can degrade by tens of percent solely by virtue  of more poloidal magnetic field orientations, rather than a lack of symmetric magnetic field conditions. This result echoes the vertical field diagrams in \citet{Narayan_2021} and \citet{Gelles_2021_published}, in which modest inclinations cause purely vertical fields to look dramatically asymmetric. 

\section{Sensitivity to $\iota$}\label{appendix:delta_iota}
\begin{figure*}
    \centering
    \includegraphics[width=1.0\linewidth]{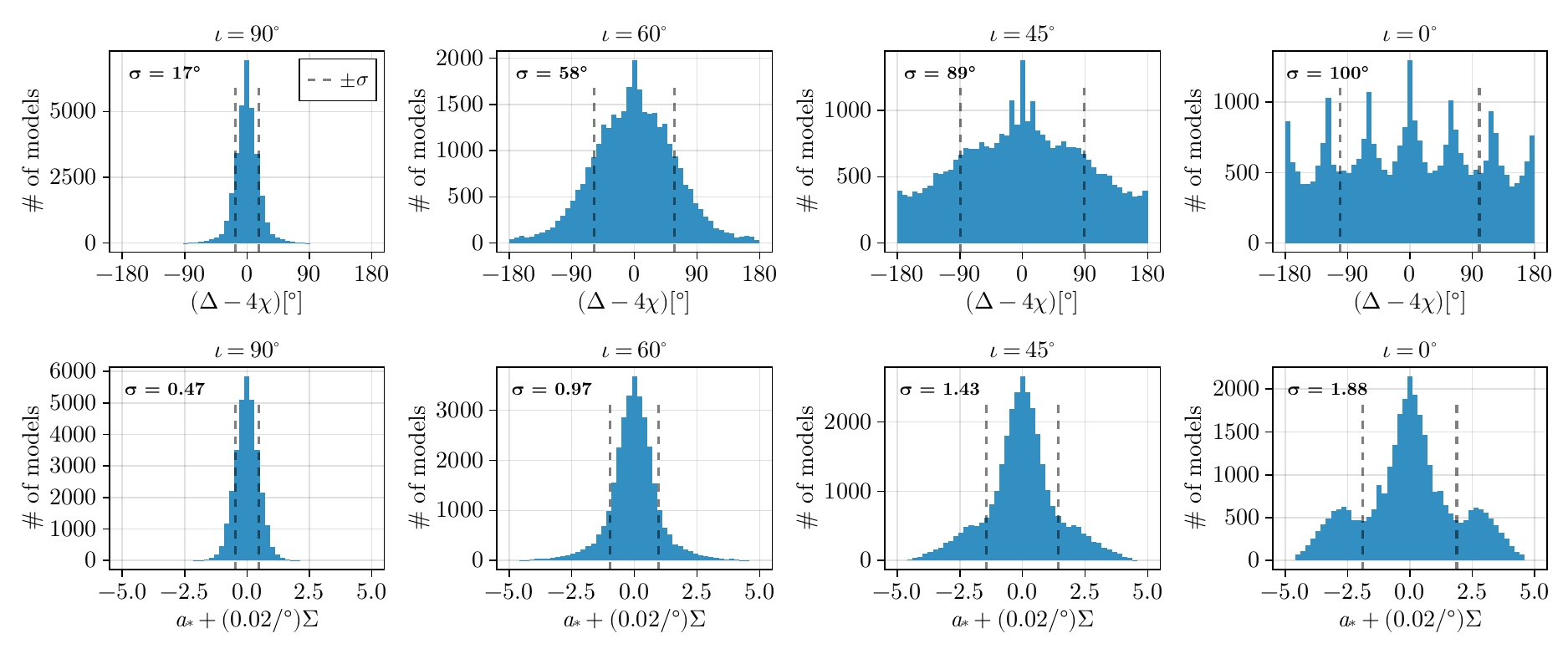}
    \caption{{Comparison of $\Delta = 4\chi$ and $a_* = -(0.02/\degree)\Sigma$ fits to models with different magnetic field inclination angles $\iota$. The top row shows the distribution of values $\Delta - 4\chi$ for different $\iota$. The bottom row shows the distribution of values $a_* + (0.02/\degree)\Sigma$. The dashed lines show the standard deviation $\sigma$ in each case. Note that, unlike in \autoref{fig:chi_constr}, there is no $\chi < -90\degree$ restriction, so the plots show distributions across all available models.}}
    \label{fig:delta_iota}
\end{figure*}

{Our main analysis finds simple relationships $\Delta = 4\chi$ and $a_* = -0.02(\Sigma/\deg)$ for models where $\iota = 90\degree$. We may study the impact of more exotic magnetic field line inclination angles by examining the distributions of the deviation from these trends as $\iota$ varies from $90\degree$. In \autoref{fig:delta_iota}, we plot the distributions of $\Delta - 4\chi$ and $a_* + 0.02(\Sigma/\deg)$ for different values of $\iota$ across our model dataset. We also indicate the corresponding standard deviations.}

{We see that decreasing the value of $\iota$ from $90\degree$ greatly increases the spread from predicted fits, making the direct inference of $\chi$ and $a_*$ from $\Delta$ and $\Sigma$ unreliable already for $\iota = 60\degree$ (at $\iota = 60\degree$, the standard deviations are $\sigma \sim 58\degree$ for $\Delta - 4\chi$, and $\sigma \sim 0.97$ for $a_* + 0.02\Sigma$). For $\iota$ of $45\degree$ or below, the distributions become clearly non-Gaussian, and take on new sub-structure as the underlying $\iota=90\degree$ \edit1{assumption} becomes \edit1{undescriptive} and the grid sampling peaks once again project out from the origin. This substructure suggests that, while being weak estimators on their own, $\Delta$ and $\Sigma$ might provide stronger constraints on $\chi$ and $a_*$ when used in combination with other observables, even in cases where the magnetic field is not aligned with the surface of the cone.}

\acknowledgments{We thank our referee for their thoughtful review of our manuscript. We are grateful to Michael Johnson for many helpful discussions. This work was conducted as part of the Black Hole Explorers program, which is supported by a generous gift by Mr. Michael and Dr. Amy Tuteur. We acknowledge financial support from the National Science Foundation (AST-2307887). This work was supported by the Black Hole Initiative, which is funded by grants from the John Templeton Foundation (Grant 62286) and the Gordon and Betty Moore Foundation (Grant GBMF-8273) - although the opinions expressed in this work are those of the author(s) and do not necessarily reflect the views of these Foundations. }

\bibliography{main}

@ARTICLE{EHT2021_2025_published,
       author = {{The Event Horizon Telescope Collaboration}},
 title = "{Horizon-scale variability of M87* from 2017─2021 EHT observations}",
      journal = {\aap},
         year = 2025,
        month = dec,
       volume = {704},
          eid = {A91},
        pages = {A91},
          doi = {10.1051/0004-6361/202555855},
archivePrefix = {arXiv},
       eprint = {2509.24593},
 primaryClass = {astro-ph.HE},
       adsurl = {https://ui.adsabs.harvard.edu/abs/2025A&A...704A..91E}
}

@ARTICLE{Chang_2024,
       author = {{Chang}, Dominic O. and {Johnson}, Michael D. and {Tiede}, Paul and {Palumbo}, Daniel C.~M.},
        title = "{Bayesian Black Hole Photogrammetry}",
      journal = {\apj},
         year = 2024,
        month = oct,
       volume = {974},
       number = {1},
          eid = {143},
        pages = {143},
          doi = {10.3847/1538-4357/ad6b28},
       adsurl = {https://ui.adsabs.harvard.edu/abs/2024ApJ...974..143C}
}

@ARTICLE{SgrA_Paper_VII,
       author = {{Event Horizon Telescope Collaboration} and {Akiyama}, Kazunori and {Alberdi}, Antxon and {Alef}, Walter and {Algaba}, Juan Carlos and {Anantua}, Richard and {Asada}, Keiichi and {Azulay}, Rebecca and {Bach}, Uwe and {Baczko}, Anne-Kathrin and {Ball}, David and {Balokovic}, Mislav and {Bandyopadhyay}, Bidisha and {Barrett}, John and {Baub{\"o}ck}, Michi and {Benson}, Bradford A. and {Bintley}, Dan and {Blackburn}, Lindy and {Blundell}, Raymond and {Bouman}, Katherine L. and {Bower}, Geoffrey C. and {Boyce}, Hope and {Bremer}, Michael and {Brinkerink}, Christiaan D. and {Brissenden}, Roger and {Britzen}, Silke and {Broderick}, Avery E. and {Broguiere}, Dominique and {Bronzwaer}, Thomas and {Bustamante}, Sandra and {Byun}, Do-Young and {Carlstrom}, John E. and {Ceccobello}, Chiara and {Chael}, Andrew and {Chan}, Chi-kwan and {Chang}, Dominic O. and {Chatterjee}, Koushik and {Chatterjee}, Shami and {Chen}, Ming-Tang and {Chen}, Yongjun and {Cheng}, Xiaopeng and {Cho}, Ilje and {Christian}, Pierre and {Conroy}, Nicholas S. and {Conway}, John E. and {Cordes}, James M. and {Crawford}, Thomas M. and {Crew}, Geoffrey B. and {Cruz-Osorio}, Alejandro and {Cui}, Yuzhu and {Dahale}, Rohan and {Davelaar}, Jordy and {De Laurentis}, Mariafelicia and {Deane}, Roger and {Dempsey}, Jessica and {Desvignes}, Gregory and {Dexter}, Jason and {Dhruv}, Vedant and {Dihingia}, Indu K. and {Doeleman}, Sheperd S. and {Dougal}, Sean Taylor and {Dzib}, Sergio A. and {Eatough}, Ralph P. and {Emami}, Razieh and {Falcke}, Heino and {Farah}, Joseph and {Fish}, Vincent L. and {Fomalont}, Edward and {Ford}, H. Alyson and {Foschi}, Marianna and {Fraga-Encinas}, Raquel and {Freeman}, William T. and {Friberg}, Per and {Fromm}, Christian M. and {Fuentes}, Antonio and {Galison}, Peter and {Gammie}, Charles F. and {Garc{\'\i}a}, Roberto and {Gentaz}, Olivier and {Georgiev}, Boris and {Goddi}, Ciriaco and {Gold}, Roman and {G{\'o}mez-Ruiz}, Arturo I. and {G{\'o}mez}, Jos{\'e} L. and {Gu}, Minfeng and {Gurwell}, Mark and {Hada}, Kazuhiro and {Haggard}, Daryl and {Haworth}, Kari and {Hecht}, Michael H. and {Hesper}, Ronald and {Heumann}, Dirk and {Ho}, Luis C. and {Ho}, Paul and {Honma}, Mareki and {Huang}, Chih-Wei L. and {Huang}, Lei and {Hughes}, David H. and {Ikeda}, Shiro and {Impellizzeri}, C.~M. Violette and {Inoue}, Makoto and {Issaoun}, Sara and {James}, David J. and {Jannuzi}, Buell T. and {Janssen}, Michael and {Jeter}, Britton and {Jiang}, Wu and {Jim{\'e}nez-Rosales}, Alejandra and {Johnson}, Michael D. and {Jorstad}, Svetlana and {Joshi}, Abhishek V. and {Jung}, Taehyun and {Karami}, Mansour and {Karuppusamy}, Ramesh and {Kawashima}, Tomohisa and {Keating}, Garrett K. and {Kettenis}, Mark and {Kim}, Dong-Jin and {Kim}, Jae-Young and {Kim}, Jongsoo and {Kim}, Junhan and {Kino}, Motoki and {Koay}, Jun Yi and {Kocherlakota}, Prashant and {Kofuji}, Yutaro and {Koch}, Patrick M. and {Koyama}, Shoko and {Kramer}, Carsten and {Kramer}, Joana A. and {Kramer}, Michael and {Krichbaum}, Thomas P. and {Kuo}, Cheng-Yu and {La Bella}, Noemi and {Lauer}, Tod R. and {Lee}, Daeyoung and {Lee}, Sang-Sung and {Leung}, Po Kin and {Levis}, Aviad and {Li}, Zhiyuan and {Lico}, Rocco and {Lindahl}, Greg and {Lindqvist}, Michael and {Lisakov}, Mikhail and {Liu}, Jun and {Liu}, Kuo and {Liuzzo}, Elisabetta and {Lo}, Wen-Ping and {Lobanov}, Andrei P. and {Loinard}, Laurent and {Lonsdale}, Colin J. and {Lowitz}, Amy E. and {Lu}, Ru-Sen and {MacDonald}, Nicholas R. and {Mao}, Jirong and {Marchili}, Nicola and {Markoff}, Sera and {Marrone}, Daniel P. and {Marscher}, Alan P. and {Mart{\'\i}-Vidal}, Iv{\'a}n and {Matsushita}, Satoki and {Matthews}, Lynn D. and {Medeiros}, Lia and {Menten}, Karl M. and {Michalik}, Daniel and {Mizuno}, Izumi and {Mizuno}, Yosuke and {Moran}, James M. and {Moriyama}, Kotaro and {Moscibrodzka}, Monika and {Mulaudzi}, Wanga and {M{\"u}ller}, Cornelia and {M{\"u}ller}, Hendrik and {Mus}, Alejandro and {Musoke}, Gibwa and {Myserlis}, Ioannis and {Nadolski}, Andrew and {Nagai}, Hiroshi and {Nagar}, Neil M. and {Nakamura}, Masanori and {Narayanan}, Gopal and {Natarajan}, Iniyan and {Nathanail}, Antonios and {Fuentes}, Santiago Navarro and {Neilsen}, Joey and {Neri}, Roberto and {Ni}, Chunchong and {Noutsos}, Aristeidis and {Nowak}, Michael A. and {Oh}, Junghwan and {Okino}, Hiroki and {Olivares}, H{\`e}ctor and {Ortiz-Le{\'o}n}, Gisela N. and {Oyama}, Tomoaki and {{\"O}zel}, Feryal and {Palumbo}, Daniel C.~M. and {Paraschos}, Georgios Filippos and {Park}, Jongho and {Parsons}, Harriet and {Patel}, Nimesh and {Pen}, Ue-Li and {Pesce}, Dominic W. and {Pi{\'e}tu}, Vincent and {Plambeck}, Richard and {PopStefanija}, Aleksandar and {Porth}, Oliver and {P{\"o}tzl}, Felix M. and {Prather}, Ben and {Preciado-L{\'o}pez}, Jorge A. and {Psaltis}, Dimitrios and {Pu}, Hung-Yi and {Ramakrishnan}, Venkatessh and {Rao}, Ramprasad and {Rawlings}, Mark G. and {Raymond}, Alexander W. and {Rezzolla}, Luciano and {Ricarte}, Angelo and {Ripperda}, Bart and {Roelofs}, Freek and {Rogers}, Alan and {Romero-Ca{\~n}izales}, Cristina and {Ros}, Eduardo and {Roshanineshat}, Arash and {Rottmann}, Helge and {Roy}, Alan L. and {Ruiz}, Ignacio and {Ruszczyk}, Chet and {Rygl}, Kazi L.~J. and {S{\'a}nchez}, Salvador and {S{\'a}nchez-Arg{\"u}elles}, David and {S{\'a}nchez-Portal}, Miguel and {Sasada}, Mahito and {Satapathy}, Kaushik and {Savolainen}, Tuomas and {Schloerb}, F. Peter and {Schonfeld}, Jonathan and {Schuster}, Karl-Friedrich and {Shao}, Lijing and {Shen}, Zhiqiang and {Small}, Des and {Sohn}, Bong Won and {SooHoo}, Jason and {Sosapanta Salas}, Le{\'o}n David and {Souccar}, Kamal and {Stanway}, Joshua S. and {Sun}, He and {Tazaki}, Fumie and {Tetarenko}, Alexandra J. and {Tiede}, Paul and {Tilanus}, Remo P.~J. and {Titus}, Michael and {Torne}, Pablo and {Toscano}, Teresa and {Traianou}, Efthalia and {Trent}, Tyler and {Trippe}, Sascha and {Turk}, Matthew and {van Bemmel}, Ilse and {van Langevelde}, Huib Jan and {van Rossum}, Daniel R. and {Vos}, Jesse and {Wagner}, Jan and {Ward-Thompson}, Derek and {Wardle}, John and {Washington}, Jasmin E. and {Weintroub}, Jonathan and {Wharton}, Robert and {Wielgus}, Maciek and {Wiik}, Kaj and {Witzel}, Gunther and {Wondrak}, Michael F. and {Wong}, George N. and {Wu}, Qingwen and {Yadlapalli}, Nitika and {Yamaguchi}, Paul and {Yfantis}, Aristomenis and {Yoon}, Doosoo and {Young}, Andr{\`e} and {Young}, Ken and {Younsi}, Ziri and {Yu}, Wei and {Yuan}, Feng and {Yuan}, Ye-Fei and {Zensus}, J. Anton and {Zhang}, Shuo and {Zhao}, Guang-Yao and {Zhao}, Shan-Shan},
        title = "{First Sagittarius A* Event Horizon Telescope Results. VII. Polarization of the Ring}",
      journal = {\apjl},
         year = 2024,
        month = apr,
       volume = {964},
       number = {2},
          eid = {L25},
        pages = {L25},
          doi = {10.3847/2041-8213/ad2df0},
       adsurl = {https://ui.adsabs.harvard.edu/abs/2024ApJ...964L..25E}
}

@ARTICLE{SgrA_Paper_VIII,
       author = {{Event Horizon Telescope Collaboration} and {Akiyama}, Kazunori and {Alberdi}, Antxon and {Alef}, Walter and {Algaba}, Juan Carlos and {Anantua}, Richard and {Asada}, Keiichi and {Azulay}, Rebecca and {Bach}, Uwe and {Baczko}, Anne-Kathrin and {Ball}, David and {Balokovi{\'c}}, Mislav and {Bandyopadhyay}, Bidisha and {Barrett}, John and {Baub{\"o}ck}, Michi and {Benson}, Bradford A. and {Bintley}, Dan and {Blackburn}, Lindy and {Blundell}, Raymond and {Bouman}, Katherine L. and {Bower}, Geoffrey C. and {Boyce}, Hope and {Bremer}, Michael and {Brinkerink}, Christiaan D. and {Brissenden}, Roger and {Britzen}, Silke and {Broderick}, Avery E. and {Broguiere}, Dominique and {Bronzwaer}, Thomas and {Bustamante}, Sandra and {Byun}, Do-Young and {Carlstrom}, John E. and {Ceccobello}, Chiara and {Chael}, Andrew and {Chan}, Chi-kwan and {Chang}, Dominic O. and {Chatterjee}, Koushik and {Chatterjee}, Shami and {Chen}, Ming-Tang and {Chen}, Yongjun and {Cheng}, Xiaopeng and {Cho}, Ilje and {Christian}, Pierre and {Conroy}, Nicholas S. and {Conway}, John E. and {Cordes}, James M. and {Crawford}, Thomas M. and {Crew}, Geoffrey B. and {Cruz-Osorio}, Alejandro and {Cui}, Yuzhu and {Dahale}, Rohan and {Davelaar}, Jordy and {De Laurentis}, Mariafelicia and {Deane}, Roger and {Dempsey}, Jessica and {Desvignes}, Gregory and {Dexter}, Jason and {Dhruv}, Vedant and {Dihingia}, Indu K. and {Doeleman}, Sheperd S. and {Dougall}, Sean and {Dzib}, Sergio A. and {Eatough}, Ralph P. and {Emami}, Razieh and {Falcke}, Heino and {Farah}, Joseph and {Fish}, Vincent L. and {Fomalont}, Edward and {Ford}, H. Alyson and {Foschi}, Marianna and {Fraga-Encinas}, Raquel and {Freeman}, William T. and {Friberg}, Per and {Fromm}, Christian M. and {Fuentes}, Antonio and {Galison}, Peter and {Gammie}, Charles F. and {Garc{\'\i}a}, Roberto and {Gentaz}, Olivier and {Georgiev}, Boris and {Goddi}, Ciriaco and {Gold}, Roman and {G{\'o}mez-Ruiz}, Arturo I. and {G{\'o}mez}, Jos{\'e} L. and {Gu}, Minfeng and {Gurwell}, Mark and {Hada}, Kazuhiro and {Haggard}, Daryl and {Haworth}, Kari and {Hecht}, Michael H. and {Hesper}, Ronald and {Heumann}, Dirk and {Ho}, Luis C. and {Ho}, Paul and {Honma}, Mareki and {Huang}, Chih-Wei L. and {Huang}, Lei and {Hughes}, David H. and {Ikeda}, Shiro and {Impellizzeri}, C.~M. Violette and {Inoue}, Makoto and {Issaoun}, Sara and {James}, David J. and {Jannuzi}, Buell T. and {Janssen}, Michael and {Jeter}, Britton and {Jiang}, Wu and {Jim{\'e}nez-Rosales}, Alejandra and {Johnson}, Michael D. and {Jorstad}, Svetlana and {Joshi}, Abhishek V. and {Jung}, Taehyun and {Karami}, Mansour and {Karuppusamy}, Ramesh and {Kawashima}, Tomohisa and {Keating}, Garrett K. and {Kettenis}, Mark and {Kim}, Dong-Jin and {Kim}, Jae-Young and {Kim}, Jongsoo and {Kim}, Junhan and {Kino}, Motoki and {Koay}, Jun Yi and {Kocherlakota}, Prashant and {Kofuji}, Yutaro and {Koch}, Patrick M. and {Koyama}, Shoko and {Kramer}, Carsten and {Kramer}, Joana A. and {Kramer}, Michael and {Krichbaum}, Thomas P. and {Kuo}, Cheng-Yu and {La Bella}, Noemi and {Lauer}, Tod R. and {Lee}, Daeyoung and {Lee}, Sang-Sung and {Leung}, Po Kin and {Levis}, Aviad and {Li}, Zhiyuan and {Lico}, Rocco and {Lindahl}, Greg and {Lindqvist}, Michael and {Lisakov}, Mikhail and {Liu}, Jun and {Liu}, Kuo and {Liuzzo}, Elisabetta and {Lo}, Wen-Ping and {Lobanov}, Andrei P. and {Loinard}, Laurent and {Lonsdale}, Colin J. and {Lowitz}, Amy E. and {Lu}, Ru-Sen and {MacDonald}, Nicholas R. and {Mao}, Jirong and {Marchili}, Nicola and {Markoff}, Sera and {Marrone}, Daniel P. and {Marscher}, Alan P. and {Mart{\'\i}-Vidal}, Iv{\'a}n and {Matsushita}, Satoki and {Matthews}, Lynn D. and {Medeiros}, Lia and {Menten}, Karl M. and {Michalik}, Daniel and {Mizuno}, Izumi and {Mizuno}, Yosuke and {Moran}, James M. and {Moriyama}, Kotaro and {Moscibrodzka}, Monika and {Mulaudzi}, Wanga and {M{\"u}ller}, Cornelia and {M{\"u}ller}, Hendrik and {Mus}, Alejandro and {Musoke}, Gibwa and {Myserlis}, Ioannis and {Nadolski}, Andrew and {Nagai}, Hiroshi and {Nagar}, Neil M. and {Nakamura}, Masanori and {Narayanan}, Gopal and {Natarajan}, Iniyan and {Nathanail}, Antonios and {Fuentes}, Santiago Navarro and {Neilsen}, Joey and {Neri}, Roberto and {Ni}, Chunchong and {Noutsos}, Aristeidis and {Nowak}, Michael A. and {Oh}, Junghwan and {Okino}, Hiroki and {Olivares}, H{\'e}ctor and {Ortiz-Le{\'o}n}, Gisela N. and {Oyama}, Tomoaki and {{\"O}zel}, Feryal and {Palumbo}, Daniel C.~M. and {Paraschos}, Georgios Filippos and {Park}, Jongho and {Parsons}, Harriet and {Patel}, Nimesh and {Pen}, Ue-Li and {Pesce}, Dominic W. and {Pi{\'e}tu}, Vincent and {Plambeck}, Richard and {PopStefanija}, Aleksandar and {Porth}, Oliver and {P{\"o}tzl}, Felix M. and {Prather}, Ben and {Preciado-L{\'o}pez}, Jorge A. and {Psaltis}, Dimitrios and {Pu}, Hung-Yi and {Ramakrishnan}, Venkatessh and {Rao}, Ramprasad and {Rawlings}, Mark G. and {Raymond}, Alexander W. and {Rezzolla}, Luciano and {Ricarte}, Angelo and {Ripperda}, Bart and {Roelofs}, Freek and {Rogers}, Alan and {Romero-Ca{\~n}izales}, Cristina and {Ros}, Eduardo and {Roshanineshat}, Arash and {Rottmann}, Helge and {Roy}, Alan L. and {Ruiz}, Ignacio and {Ruszczyk}, Chet and {Rygl}, Kazi L.~J. and {S{\'a}nchez}, Salvador and {S{\'a}nchez-Arg{\"u}elles}, David and {S{\'a}nchez-Portal}, Miguel and {Sasada}, Mahito and {Satapathy}, Kaushik and {Savolainen}, Tuomas and {Schloerb}, F. Peter and {Schonfeld}, Jonathan and {Schuster}, Karl-Friedrich and {Shao}, Lijing and {Shen}, Zhiqiang and {Small}, Des and {Sohn}, Bong Won and {SooHoo}, Jason and {Sosapanta Salas}, Le{\'o}n David and {Souccar}, Kamal and {Stanway}, Joshua S. and {Sun}, He and {Tazaki}, Fumie and {Tetarenko}, Alexandra J. and {Tiede}, Paul and {Tilanus}, Remo P.~J. and {Titus}, Michael and {Torne}, Pablo and {Toscano}, Teresa and {Traianou}, Efthalia and {Trent}, Tyler and {Trippe}, Sascha and {Turk}, Matthew and {van Bemmel}, Ilse and {van Langevelde}, Huib Jan and {van Rossum}, Daniel R. and {Vos}, Jesse and {Wagner}, Jan and {Ward-Thompson}, Derek and {Wardle}, John and {Washington}, Jasmin E. and {Weintroub}, Jonathan and {Wharton}, Robert and {Wielgus}, Maciek and {Wiik}, Kaj and {Witzel}, Gunther and {Wondrak}, Michael F. and {Wong}, George N. and {Wu}, Qingwen and {Yadlapalli}, Nitika and {Yamaguchi}, Paul and {Yfantis}, Aristomenis and {Yoon}, Doosoo and {Young}, Andr{\'e} and {Young}, Ken and {Younsi}, Ziri and {Yu}, Wei and {Yuan}, Feng and {Yuan}, Ye-Fei and {Zensus}, J. Anton and {Zhang}, Shuo and {Zhao}, Guang-Yao and {Zhao}, Shan-Shan and {Najafi-Ziyazi}, Mahdi},
        title = "{First Sagittarius A* Event Horizon Telescope Results. VIII. Physical Interpretation of the Polarized Ring}",
      journal = {\apjl},
         year = 2024,
        month = apr,
       volume = {964},
       number = {2},
          eid = {L26},
        pages = {L26},
          doi = {10.3847/2041-8213/ad2df1},
       adsurl = {https://ui.adsabs.harvard.edu/abs/2024ApJ...964L..26E}
}

@ARTICLE{PWP_2020,
       author = {{Palumbo}, Daniel C.~M. and {Wong}, George N. and {Prather}, Ben S.},
        title = "{Discriminating Accretion States via Rotational Symmetry in Simulated Polarimetric Images of M87}",
      journal = {\apj},
         year = 2020,
        month = may,
       volume = {894},
       number = {2},
          eid = {156},
        pages = {156},
          doi = {10.3847/1538-4357/ab86ac},
archivePrefix = {arXiv},
       eprint = {2004.01751},
 primaryClass = {astro-ph.HE},
       adsurl = {https://ui.adsabs.harvard.edu/abs/2020ApJ...894..156P}
}

@ARTICLE{Palumbo_2022,
       author = {{Palumbo}, Daniel C.~M. and {Wong}, George N.},
        title = "{Photon Ring Symmetries in Simulated Linear Polarization Images of Messier 87*}",
      journal = {\apj},
         year = 2022,
        month = apr,
       volume = {929},
       number = {1},
          eid = {49},
        pages = {49},
          doi = {10.3847/1538-4357/ac59b4},
archivePrefix = {arXiv},
       eprint = {2203.00844},
 primaryClass = {astro-ph.HE},
       adsurl = {https://ui.adsabs.harvard.edu/abs/2022ApJ...929...49P}
}

@ARTICLE{Chael_2023,
       author = {{Chael}, Andrew and {Lupsasca}, Alexandru and {Wong}, George N. and {Quataert}, Eliot},
        title = "{Black Hole Polarimetry I. A Signature of Electromagnetic Energy Extraction}",
      journal = {\apj},
         year = 2023,
        month = nov,
       volume = {958},
       number = {1},
          eid = {65},
        pages = {65},
          doi = {10.3847/1538-4357/acf92d},
archivePrefix = {arXiv},
       eprint = {2307.06372},
 primaryClass = {astro-ph.HE},
       adsurl = {https://ui.adsabs.harvard.edu/abs/2023ApJ...958...65C}
}

@ARTICLE{Tamar_2024_published,
       author = {{Tamar}, Aditya and {Palumbo}, Daniel C.~M.},
        title = "{Photon Ring Polarimetry with Next-generation Black Hole Imaging. I. M87*}",
      journal = {\apj},
         year = 2024,
        month = dec,
       volume = {977},
       number = {2},
          eid = {147},
        pages = {147},
          doi = {10.3847/1538-4357/ad8dd5},
archivePrefix = {arXiv},
       eprint = {2410.15325},
 primaryClass = {astro-ph.HE},
       adsurl = {https://ui.adsabs.harvard.edu/abs/2024ApJ...977..147T}
}

@ARTICLE{Deglinnocenti_1985,
       author = {{Landi Degl'Innocenti}, E. and {Landi Degl'Innocenti}, M.},
        title = "{On the solution of the radiative transfer equations for polarized radiation}",
      journal = {\solphys},
         year = 1985,
        month = jun,
       volume = {97},
        pages = {239-250},
          doi = {10.1007/BF00165988},
       adsurl = {https://ui.adsabs.harvard.edu/abs/1985SoPh...97..239L}
}

@ARTICLE{Palumbo_KerrBAM,
       author = {{Palumbo}, Daniel C.~M. and {Gelles}, Zachary and {Tiede}, Paul and {Chang}, Dominic O. and {Pesce}, Dominic W. and {Chael}, Andrew and {Johnson}, Michael D.},
        title = "{Bayesian Accretion Modeling: Axisymmetric Equatorial Emission in the Kerr Spacetime}",
      journal = {\apj},
         year = 2022,
        month = nov,
       volume = {939},
       number = {2},
          eid = {107},
        pages = {107},
          doi = {10.3847/1538-4357/ac9ab7},
archivePrefix = {arXiv},
       eprint = {2210.07108},
 primaryClass = {astro-ph.HE},
       adsurl = {https://ui.adsabs.harvard.edu/abs/2022ApJ...939..107P}
}

@ARTICLE{PaperI,
   author = {{Event Horizon Telescope Collaboration} and {Akiyama}, K. and 
	{Alberdi}, A. and {Alef}, W. and {Asada}, K. and {Azulay}, R. and 
	{Baczko}, A.-K. and {Ball}, D. and {Balokovi{\'c}}, M. and {Barrett}, J. and et al.},
    title = "{First M87 Event Horizon Telescope Results. I. The Shadow of the Supermassive Black Hole}",
  journal = {\apjl},
archivePrefix = "arXiv",
   eprint = {1906.11238},
     year = "{2019a}",
    month = apr,
   volume = 875,
      eid = {L1},
    pages = {L1},
      doi = {10.3847/2041-8213/ab0ec7},
   adsurl = {https://ui.adsabs.harvard.edu/abs/2019ApJ...875L...1E}
}

@ARTICLE{PaperII,
   author = {{Event Horizon Telescope Collaboration} and {Akiyama}, K. and 
	{Alberdi}, A. and {Alef}, W. and {Asada}, K. and {Azulay}, R. and 
	{Baczko}, A.-K. and {Ball}, D. and {Balokovi{\'c}}, M. and {Barrett}, J. and et al.},
    title = "{First M87 Event Horizon Telescope Results. II. Array and Instrumentation}",
  journal = {\apjl},
archivePrefix = "arXiv",
   eprint = {1906.11239},
 primaryClass = "astro-ph.IM",
     year = "{2019b}",
    month = apr,
   volume = 875,
      eid = {L2},
    pages = {L2},
      doi = {10.3847/2041-8213/ab0c96},
   adsurl = {https://ui.adsabs.harvard.edu/abs/2019ApJ...875L...2E}
}

@ARTICLE{PaperIII,
   author = {{Event Horizon Telescope Collaboration} and {Akiyama}, K. and 
	{Alberdi}, A. and {Alef}, W. and {Asada}, K. and {Azulay}, R. and 
	{Baczko}, A.-K. and {Ball}, D. and {Balokovi{\'c}}, M. and {Barrett}, J. and et al.},
    title = "{First M87 Event Horizon Telescope Results. III. Data Processing and Calibration}",
  journal = {\apjl},
     year = "{2019c}",
    month = apr,
   volume = 875,
      eid = {L3},
    pages = {L3},
      doi = {10.3847/2041-8213/ab0c57},
   adsurl = {http://adsabs.harvard.edu/abs/2019ApJ...875L...3E}
}

@ARTICLE{PaperIV,
   author = {{Event Horizon Telescope Collaboration} and {Akiyama}, K. and 
	{Alberdi}, A. and {Alef}, W. and {Asada}, K. and {Azulay}, R. and 
	{Baczko}, A.-K. and {Ball}, D. and {Balokovi{\'c}}, M. and {Barrett}, J. and et al.},
    title = "{First M87 Event Horizon Telescope Results. IV. Imaging the Central Supermassive Black Hole}",
  journal = {\apjl},
     year = "{2019d}",
    month = apr,
   volume = 875,
      eid = {L4},
    pages = {L4},
      doi = {10.3847/2041-8213/ab0e85},
   adsurl = {https://ui.adsabs.harvard.edu/abs/2019ApJ...875L...4E}
}

@ARTICLE{PaperV,
   author = {{Event Horizon Telescope Collaboration} and {Akiyama}, K. and 
	{Alberdi}, A. and {Alef}, W. and {Asada}, K. and {Azulay}, R. and 
	{Baczko}, A.-K. and {Ball}, D. and {Balokovi{\'c}}, M. and {Barrett}, J. and et al.},
    title = "{First M87 Event Horizon Telescope Results. V. Physical Origin of the Asymmetric Ring}",
  journal = {\apjl},
archivePrefix = "arXiv",
   eprint = {1906.11242},
     year = "{2019e}",
    month = apr,
   volume = 875,
      eid = {L5},
    pages = {L5},
      doi = {10.3847/2041-8213/ab0f43},
   adsurl = {https://ui.adsabs.harvard.edu/abs/2019ApJ...875L...5E}
}

@ARTICLE{PaperVI,
   author = {{Event Horizon Telescope Collaboration} and {Akiyama}, K. and 
	{Alberdi}, A. and {Alef}, W. and {Asada}, K. and {Azulay}, R. and 
	{Baczko}, A.-K. and {Ball}, D. and {Balokovi{\'c}}, M. and {Barrett}, J. and et al.},
    title = "{First M87 Event Horizon Telescope Results. VI. The Shadow and Mass of the Central Black Hole}",
  journal = {\apjl},
archivePrefix = "arXiv",
   eprint = {1906.11243},
     year = "{2019f}",
    month = apr,
   volume = 875,
      eid = {L6},
    pages = {L6},
      doi = {10.3847/2041-8213/ab1141},
   adsurl = {https://ui.adsabs.harvard.edu/abs/2019ApJ...875L...6E}
}

@ARTICLE{Doeleman_2023,
       author = {{Doeleman}, Sheperd S. and {Barrett}, John and {Blackburn}, Lindy and {Bouman}, Katherine L. and {Broderick}, Avery E. and {Chaves}, Ryan and {Fish}, Vincent L. and {Fitzpatrick}, Garret and {Freeman}, Mark and {Fuentes}, Antonio and {G{\'o}mez}, Jos{\'e} L. and {Haworth}, Kari and {Houston}, Janice and {Issaoun}, Sara and {Johnson}, Michael D. and {Kettenis}, Mark and {Loinard}, Laurent and {Nagar}, Neil and {Narayanan}, Gopal and {Oppenheimer}, Aaron and {Palumbo}, Daniel C.~M. and {Patel}, Nimesh and {Pesce}, Dominic W. and {Raymond}, Alexander W. and {Roelofs}, Freek and {Srinivasan}, Ranjani and {Tiede}, Paul and {Weintroub}, Jonathan and {Wielgus}, Maciek},
        title = "{Reference Array and Design Consideration for the Next-Generation Event Horizon Telescope}",
      journal = {Galaxies},
         year = 2023,
        month = oct,
       volume = {11},
       number = {5},
          eid = {107},
        pages = {107},
          doi = {10.3390/galaxies11050107},
archivePrefix = {arXiv},
       eprint = {2306.08787},
 primaryClass = {astro-ph.IM},
       adsurl = {https://ui.adsabs.harvard.edu/abs/2023Galax..11..107D}
}

@ARTICLE{Emami_2023,
       author = {{Emami}, Razieh and {Ricarte}, Angelo and {Wong}, George N. and {Palumbo}, Daniel and {Chang}, Dominic and {Doeleman}, Sheperd S. and {Broderick}, Avery E. and {Narayan}, Ramesh and {Wielgus}, Maciek and {Blackburn}, Lindy and {Prather}, Ben S. and {Chael}, Andrew A. and {Anantua}, Richard and {Chatterjee}, Koushik and {Marti-Vidal}, Ivan and {G{\'o}mez}, Jose L. and {Akiyama}, Kazunori and {Liska}, Matthew and {Hernquist}, Lars and {Tremblay}, Grant and {Vogelsberger}, Mark and {Alcock}, Charles and {Smith}, Randall and {Steiner}, James and {Tiede}, Paul and {Roelofs}, Freek},
        title = "{Unraveling Twisty Linear Polarization Morphologies in Black Hole Images}",
      journal = {\apj},
         year = 2023,
        month = jun,
       volume = {950},
       number = {1},
          eid = {38},
        pages = {38},
          doi = {10.3847/1538-4357/acc8cd},
archivePrefix = {arXiv},
       eprint = {2210.01218},
 primaryClass = {astro-ph.GA},
       adsurl = {https://ui.adsabs.harvard.edu/abs/2023ApJ...950...38E}
}

@ARTICLE{Johnson_2020,
       author = {{Johnson}, Michael D. and {Lupsasca}, Alexandru and {Strominger}, Andrew and {Wong}, George N. and {Hadar}, Shahar and {Kapec}, Daniel and {Narayan}, Ramesh and {Chael}, Andrew and {Gammie}, Charles F. and {Galison}, Peter and {Palumbo}, Daniel C.~M. and {Doeleman}, Sheperd S. and {Blackburn}, Lindy and {Wielgus}, Maciek and {Pesce}, Dominic W. and {Farah}, Joseph R. and {Moran}, James M.},
        title = "{Universal interferometric signatures of a black hole's photon ring}",
      journal = {Science Advances},
         year = 2020,
        month = mar,
       volume = {6},
       number = {12},
        pages = {eaaz1310},
          doi = {10.1126/sciadv.aaz1310},
archivePrefix = {arXiv},
       eprint = {1907.04329},
 primaryClass = {astro-ph.IM},
       adsurl = {https://ui.adsabs.harvard.edu/abs/2020SciA....6.1310J}
}

@ARTICLE{PaperVII,
       author = {{Event Horizon Telescope Collaboration} and {Akiyama}, Kazunori and {Algaba}, Juan Carlos and {Alberdi}, Antxon and {Alef}, Walter and {Anantua}, Richard and {Asada}, Keiichi and {Azulay}, Rebecca and {Baczko}, Anne-Kathrin and {Ball}, David and {Balokovi{\'c}}, Mislav and {Barrett}, John and {Benson}, Bradford A. and {Bintley}, Dan and {Blackburn}, Lindy and {Blundell}, Raymond and {Boland}, Wilfred and {Bouman}, Katherine L. and {Bower}, Geoffrey C. and {Boyce}, Hope and {Bremer}, Michael and {Brinkerink}, Christiaan D. and {Brissenden}, Roger and {Britzen}, Silke and {Broderick}, Avery E. and {Broguiere}, Dominique and {Bronzwaer}, Thomas and {Byun}, Do-Young and {Carlstrom}, John E. and {Chael}, Andrew and {Chan}, Chi-kwan and {Chatterjee}, Shami and {Chatterjee}, Koushik and {Chen}, Ming-Tang and {Chen}, Yongjun and {Chesler}, Paul M. and {Cho}, Ilje and {Christian}, Pierre and {Conway}, John E. and {Cordes}, James M. and {Crawford}, Thomas M. and {Crew}, Geoffrey B. and {Cruz-Osorio}, Alejandro and {Cui}, Yuzhu and {Davelaar}, Jordy and {De Laurentis}, Mariafelicia and {Deane}, Roger and {Dempsey}, Jessica and {Desvignes}, Gregory and {Dexter}, Jason and {Doeleman}, Sheperd S. and {Eatough}, Ralph P. and {Falcke}, Heino and {Farah}, Joseph and {Fish}, Vincent L. and {Fomalont}, Ed and {Ford}, H. Alyson and {Fraga-Encinas}, Raquel and {Freeman}, William T. and {Friberg}, Per and {Fromm}, Christian M. and {Fuentes}, Antonio and {Galison}, Peter and {Gammie}, Charles F. and {Garc{\'\i}a}, Roberto and {Gentaz}, Olivier and {Georgiev}, Boris and {Goddi}, Ciriaco and {Gold}, Roman and {G{\'o}mez}, Jos{\'e} L. and {G{\'o}mez-Ruiz}, Arturo I. and {Gu}, Minfeng and {Gurwell}, Mark and {Hada}, Kazuhiro and {Haggard}, Daryl and {Hecht}, Michael H. and {Hesper}, Ronald and {Ho}, Luis C. and {Ho}, Paul and {Honma}, Mareki and {Huang}, Chih-Wei L. and {Huang}, Lei and {Hughes}, David H. and {Ikeda}, Shiro and {Inoue}, Makoto and {Issaoun}, Sara and {James}, David J. and {Jannuzi}, Buell T. and {Janssen}, Michael and {Jeter}, Britton and {Jiang}, Wu and {Jimenez-Rosales}, Alejandra and {Johnson}, Michael D. and {Jorstad}, Svetlana and {Jung}, Taehyun and {Karami}, Mansour and {Karuppusamy}, Ramesh and {Kawashima}, Tomohisa and {Keating}, Garrett K. and {Kettenis}, Mark and {Kim}, Dong-Jin and {Kim}, Jae-Young and {Kim}, Jongsoo and {Kim}, Junhan and {Kino}, Motoki and {Koay}, Jun Yi and {Kofuji}, Yutaro and {Koch}, Patrick M. and {Koyama}, Shoko and {Kramer}, Michael and {Kramer}, Carsten and {Krichbaum}, Thomas P. and {Kuo}, Cheng-Yu and {Lauer}, Tod R. and {Lee}, Sang-Sung and {Levis}, Aviad and {Li}, Yan-Rong and {Li}, Zhiyuan and {Lindqvist}, Michael and {Lico}, Rocco and {Lindahl}, Greg and {Liu}, Jun and {Liu}, Kuo and {Liuzzo}, Elisabetta and {Lo}, Wen-Ping and {Lobanov}, Andrei P. and {Loinard}, Laurent and {Lonsdale}, Colin and {Lu}, Ru-Sen and {MacDonald}, Nicholas R. and {Mao}, Jirong and {Marchili}, Nicola and {Markoff}, Sera and {Marrone}, Daniel P. and {Marscher}, Alan P. and {Mart{\'\i}-Vidal}, Iv{\'a}n and {Matsushita}, Satoki and {Matthews}, Lynn D. and {Medeiros}, Lia and {Menten}, Karl M. and {Mizuno}, Izumi and {Mizuno}, Yosuke and {Moran}, James M. and {Moriyama}, Kotaro and {Moscibrodzka}, Monika and {M{\"u}ller}, Cornelia and {Musoke}, Gibwa and {Mej{\'\i}as}, Alejandro Mus and {Michalik}, Daniel and {Nadolski}, Andrew and {Nagai}, Hiroshi and {Nagar}, Neil M. and {Nakamura}, Masanori and {Narayan}, Ramesh and {Narayanan}, Gopal and {Natarajan}, Iniyan and {Nathanail}, Antonios and {Neilsen}, Joey and {Neri}, Roberto and {Ni}, Chunchong and {Noutsos}, Aristeidis and {Nowak}, Michael A. and {Okino}, Hiroki and {Olivares}, H{\'e}ctor and {Ortiz-Le{\'o}n}, Gisela N. and {Oyama}, Tomoaki and {{\"O}zel}, Feryal and {Palumbo}, Daniel C.~M. and {Park}, Jongho and {Patel}, Nimesh and {Pen}, Ue-Li and {Pesce}, Dominic W. and {Pi{\'e}tu}, Vincent and {Plambeck}, Richard and {PopStefanija}, Aleksandar and {Porth}, Oliver and {P{\"o}tzl}, Felix M. and {Prather}, Ben and {Preciado-L{\'o}pez}, Jorge A. and {Psaltis}, Dimitrios and {Pu}, Hung-Yi and {Ramakrishnan}, Venkatessh and {Rao}, Ramprasad and {Rawlings}, Mark G. and {Raymond}, Alexander W. and {Rezzolla}, Luciano and {Ricarte}, Angelo and {Ripperda}, Bart and {Roelofs}, Freek and {Rogers}, Alan and {Ros}, Eduardo and {Rose}, Mel and {Roshanineshat}, Arash and {Rottmann}, Helge and {Roy}, Alan L. and {Ruszczyk}, Chet and {Rygl}, Kazi L.~J. and {S{\'a}nchez}, Salvador and {S{\'a}nchez-Arguelles}, David and {Sasada}, Mahito and {Savolainen}, Tuomas and {Schloerb}, F. Peter and {Schuster}, Karl-Friedrich and {Shao}, Lijing and {Shen}, Zhiqiang and {Small}, Des and {Sohn}, Bong Won and {SooHoo}, Jason and {Sun}, He and {Tazaki}, Fumie and {Tetarenko}, Alexandra J. and {Tiede}, Paul and {Tilanus}, Remo P.~J. and {Titus}, Michael and {Toma}, Kenji and {Torne}, Pablo and {Trent}, Tyler and {Traianou}, Efthalia and {Trippe}, Sascha and {van Bemmel}, Ilse and {van Langevelde}, Huib Jan and {van Rossum}, Daniel R. and {Wagner}, Jan and {Ward-Thompson}, Derek and {Wardle}, John and {Weintroub}, Jonathan and {Wex}, Norbert and {Wharton}, Robert and {Wielgus}, Maciek and {Wong}, George N. and {Wu}, Qingwen and {Yoon}, Doosoo and {Young}, Andr{\'e} and {Young}, Ken and {Younsi}, Ziri and {Yuan}, Feng and {Yuan}, Ye-Fei and {Zensus}, J. Anton and {Zhao}, Guang-Yao and {Zhao}, Shan-Shan},
        title = "{First M87 Event Horizon Telescope Results. VII. Polarization of the Ring}",
      journal = {\apjl},
         year = 2021,
        month = mar,
       volume = {910},
       number = {1},
          eid = {L12},
        pages = {L12},
          doi = {10.3847/2041-8213/abe71d},
archivePrefix = {arXiv},
       eprint = {2105.01169},
 primaryClass = {astro-ph.HE},
       adsurl = {https://ui.adsabs.harvard.edu/abs/2021ApJ...910L..12E}
}

@ARTICLE{PaperVIII,
       author = {{Event Horizon Telescope Collaboration} and {Akiyama}, Kazunori and {Algaba}, Juan Carlos and {Alberdi}, Antxon and {Alef}, Walter and {Anantua}, Richard and {Asada}, Keiichi and {Azulay}, Rebecca and {Baczko}, Anne-Kathrin and {Ball}, David and {Balokovi{\'c}}, Mislav and {Barrett}, John and {Benson}, Bradford A. and {Bintley}, Dan and {Blackburn}, Lindy and {Blundell}, Raymond and {Boland}, Wilfred and {Bouman}, Katherine L. and {Bower}, Geoffrey C. and {Boyce}, Hope and {Bremer}, Michael and {Brinkerink}, Christiaan D. and {Brissenden}, Roger and {Britzen}, Silke and {Broderick}, Avery E. and {Broguiere}, Dominique and {Bronzwaer}, Thomas and {Byun}, Do-Young and {Carlstrom}, John E. and {Chael}, Andrew and {Chan}, Chi-kwan and {Chatterjee}, Shami and {Chatterjee}, Koushik and {Chen}, Ming-Tang and {Chen}, Yongjun and {Chesler}, Paul M. and {Cho}, Ilje and {Christian}, Pierre and {Conway}, John E. and {Cordes}, James M. and {Crawford}, Thomas M. and {Crew}, Geoffrey B. and {Cruz-Osorio}, Alejandro and {Cui}, Yuzhu and {Davelaar}, Jordy and {De Laurentis}, Mariafelicia and {Deane}, Roger and {Dempsey}, Jessica and {Desvignes}, Gregory and {Dexter}, Jason and {Doeleman}, Sheperd S. and {Eatough}, Ralph P. and {Falcke}, Heino and {Farah}, Joseph and {Fish}, Vincent L. and {Fomalont}, Ed and {Ford}, H. Alyson and {Fraga-Encinas}, Raquel and {Friberg}, Per and {Fromm}, Christian M. and {Fuentes}, Antonio and {Galison}, Peter and {Gammie}, Charles F. and {Garc{\'\i}a}, Roberto and {Gelles}, Zachary and {Gentaz}, Olivier and {Georgiev}, Boris and {Goddi}, Ciriaco and {Gold}, Roman and {G{\'o}mez}, Jos{\'e} L. and {G{\'o}mez-Ruiz}, Arturo I. and {Gu}, Minfeng and {Gurwell}, Mark and {Hada}, Kazuhiro and {Haggard}, Daryl and {Hecht}, Michael H. and {Hesper}, Ronald and {Himwich}, Elizabeth and {Ho}, Luis C. and {Ho}, Paul and {Honma}, Mareki and {Huang}, Chih-Wei L. and {Huang}, Lei and {Hughes}, David H. and {Ikeda}, Shiro and {Inoue}, Makoto and {Issaoun}, Sara and {James}, David J. and {Jannuzi}, Buell T. and {Janssen}, Michael and {Jeter}, Britton and {Jiang}, Wu and {Jimenez-Rosales}, Alejandra and {Johnson}, Michael D. and {Jorstad}, Svetlana and {Jung}, Taehyun and {Karami}, Mansour and {Karuppusamy}, Ramesh and {Kawashima}, Tomohisa and {Keating}, Garrett K. and {Kettenis}, Mark and {Kim}, Dong-Jin and {Kim}, Jae-Young and {Kim}, Jongsoo and {Kim}, Junhan and {Kino}, Motoki and {Koay}, Jun Yi and {Kofuji}, Yutaro and {Koch}, Patrick M. and {Koyama}, Shoko and {Kramer}, Michael and {Kramer}, Carsten and {Krichbaum}, Thomas P. and {Kuo}, Cheng-Yu and {Lauer}, Tod R. and {Lee}, Sang-Sung and {Levis}, Aviad and {Li}, Yan-Rong and {Li}, Zhiyuan and {Lindqvist}, Michael and {Lico}, Rocco and {Lindahl}, Greg and {Liu}, Jun and {Liu}, Kuo and {Liuzzo}, Elisabetta and {Lo}, Wen-Ping and {Lobanov}, Andrei P. and {Loinard}, Laurent and {Lonsdale}, Colin and {Lu}, Ru-Sen and {MacDonald}, Nicholas R. and {Mao}, Jirong and {Marchili}, Nicola and {Markoff}, Sera and {Marrone}, Daniel P. and {Marscher}, Alan P. and {Mart{\'\i}-Vidal}, Iv{\'a}n and {Matsushita}, Satoki and {Matthews}, Lynn D. and {Medeiros}, Lia and {Menten}, Karl M. and {Mizuno}, Izumi and {Mizuno}, Yosuke and {Moran}, James M. and {Moriyama}, Kotaro and {Moscibrodzka}, Monika and {M{\"u}ller}, Cornelia and {Musoke}, Gibwa and {Mus Mej{\'\i}as}, Alejandro and {Michalik}, Daniel and {Nadolski}, Andrew and {Nagai}, Hiroshi and {Nagar}, Neil M. and {Nakamura}, Masanori and {Narayan}, Ramesh and {Narayanan}, Gopal and {Natarajan}, Iniyan and {Nathanail}, Antonios and {Neilsen}, Joey and {Neri}, Roberto and {Ni}, Chunchong and {Noutsos}, Aristeidis and {Nowak}, Michael A. and {Okino}, Hiroki and {Olivares}, H{\'e}ctor and {Ortiz-Le{\'o}n}, Gisela N. and {Oyama}, Tomoaki and {{\"O}zel}, Feryal and {Palumbo}, Daniel C.~M. and {Park}, Jongho and {Patel}, Nimesh and {Pen}, Ue-Li and {Pesce}, Dominic W. and {Pi{\'e}tu}, Vincent and {Plambeck}, Richard and {PopStefanija}, Aleksandar and {Porth}, Oliver and {P{\"o}tzl}, Felix M. and {Prather}, Ben and {Preciado-L{\'o}pez}, Jorge A. and {Psaltis}, Dimitrios and {Pu}, Hung-Yi and {Ramakrishnan}, Venkatessh and {Rao}, Ramprasad and {Rawlings}, Mark G. and {Raymond}, Alexander W. and {Rezzolla}, Luciano and {Ricarte}, Angelo and {Ripperda}, Bart and {Roelofs}, Freek and {Rogers}, Alan and {Ros}, Eduardo and {Rose}, Mel and {Roshanineshat}, Arash and {Rottmann}, Helge and {Roy}, Alan L. and {Ruszczyk}, Chet and {Rygl}, Kazi L.~J. and {S{\'a}nchez}, Salvador and {S{\'a}nchez-Arguelles}, David and {Sasada}, Mahito and {Savolainen}, Tuomas and {Schloerb}, F. Peter and {Schuster}, Karl-Friedrich and {Shao}, Lijing and {Shen}, Zhiqiang and {Small}, Des and {Sohn}, Bong Won and {SooHoo}, Jason and {Sun}, He and {Tazaki}, Fumie and {Tetarenko}, Alexandra J. and {Tiede}, Paul and {Tilanus}, Remo P.~J. and {Titus}, Michael and {Toma}, Kenji and {Torne}, Pablo and {Trent}, Tyler and {Traianou}, Efthalia and {Trippe}, Sascha and {van Bemmel}, Ilse and {van Langevelde}, Huib Jan and {van Rossum}, Daniel R. and {Wagner}, Jan and {Ward-Thompson}, Derek and {Wardle}, John and {Weintroub}, Jonathan and {Wex}, Norbert and {Wharton}, Robert and {Wielgus}, Maciek and {Wong}, George N. and {Wu}, Qingwen and {Yoon}, Doosoo and {Young}, Andr{\'e} and {Young}, Ken and {Younsi}, Ziri and {Yuan}, Feng and {Yuan}, Ye-Fei and {Zensus}, J. Anton and {Zhao}, Guang-Yao and {Zhao}, Shan-Shan},
        title = "{First M87 Event Horizon Telescope Results. VIII. Magnetic Field Structure near The Event Horizon}",
      journal = {\apjl},
         year = 2021,
        month = mar,
       volume = {910},
       number = {1},
          eid = {L13},
        pages = {L13},
          doi = {10.3847/2041-8213/abe4de},
archivePrefix = {arXiv},
       eprint = {2105.01173},
 primaryClass = {astro-ph.HE},
       adsurl = {https://ui.adsabs.harvard.edu/abs/2021ApJ...910L..13E}
}

@ARTICLE{Wong_2021,
       author = {{Wong}, George N. and {Du}, Yufeng and {Prather}, Ben S. and {Gammie}, Charles F.},
        title = "{The Jet-disk Boundary Layer in Black Hole Accretion}",
      journal = {\apj},
         year = 2021,
        month = jun,
       volume = {914},
       number = {1},
          eid = {55},
        pages = {55},
          doi = {10.3847/1538-4357/abf8b8},
archivePrefix = {arXiv},
       eprint = {2104.07035},
 primaryClass = {astro-ph.HE},
       adsurl = {https://ui.adsabs.harvard.edu/abs/2021ApJ...914...55W}
}

@INPROCEEDINGS{Bardeen_1973,
       author = {{Bardeen}, J.~M.},
        title = "{Timelike and null geodesics in the Kerr metric.}",
    booktitle = {Black Holes (Les Astres Occlus)},
         year = 1973,
        month = jan,
        pages = {215-239},
       adsurl = {https://ui.adsabs.harvard.edu/abs/1973blho.conf..215B}
}

@ARTICLE{Luminet_1979,
       author = {{Luminet}, J. -P.},
        title = "{Image of a spherical black hole with thin accretion disk.}",
      journal = {\aap},
         year = 1979,
        month = may,
       volume = {75},
        pages = {228-235},
       adsurl = {https://ui.adsabs.harvard.edu/abs/1979A&A....75..228L}
}

@ARTICLE{Johannsen_2010,
       author = {{Johannsen}, Tim and {Psaltis}, Dimitrios},
        title = "{Testing the No-hair Theorem with Observations in the Electromagnetic Spectrum. II. Black Hole Images}",
      journal = {\apj},
         year = 2010,
        month = jul,
       volume = {718},
       number = {1},
        pages = {446-454},
          doi = {10.1088/0004-637X/718/1/446},
archivePrefix = {arXiv},
       eprint = {1005.1931},
 primaryClass = {astro-ph.HE},
       adsurl = {https://ui.adsabs.harvard.edu/abs/2010ApJ...718..446J}
}

@ARTICLE{SgrA_PaperIV,
       author = {{Event Horizon Telescope Collaboration} and {Akiyama}, Kazunori and {Alberdi}, Antxon and {Alef}, Walter and {Algaba}, Juan Carlos and {Anantua}, Richard and {Asada}, Keiichi and {Azulay}, Rebecca and {Bach}, Uwe and {Baczko}, Anne-Kathrin and {Ball}, David and {Balokovi{\'c}}, Mislav and {Barrett}, John and {Baub{\"o}ck}, Michi and {Benson}, Bradford A. and {Bintley}, Dan and {Blackburn}, Lindy and {Blundell}, Raymond and {Bouman}, Katherine L. and {Bower}, Geoffrey C. and {Boyce}, Hope and {Bremer}, Michael and {Brinkerink}, Christiaan D. and {Brissenden}, Roger and {Britzen}, Silke and {Broderick}, Avery E. and {Broguiere}, Dominique and {Bronzwaer}, Thomas and {Bustamante}, Sandra and {Byun}, Do-Young and {Carlstrom}, John E. and {Ceccobello}, Chiara and {Chael}, Andrew and {Chan}, Chi-kwan and {Chatterjee}, Koushik and {Chatterjee}, Shami and {Chen}, Ming-Tang and {Chen}, Yongjun and {Cheng}, Xiaopeng and {Cho}, Ilje and {Christian}, Pierre and {Conroy}, Nicholas S. and {Conway}, John E. and {Cordes}, James M. and {Crawford}, Thomas M. and {Crew}, Geoffrey B. and {Cruz-Osorio}, Alejandro and {Cui}, Yuzhu and {Davelaar}, Jordy and {Laurentis}, Mariafelicia De and {Deane}, Roger and {Dempsey}, Jessica and {Desvignes}, Gregory and {Dexter}, Jason and {Dhruv}, Vedant and {Doeleman}, Sheperd S. and {Dougal}, Sean and {Dzib}, Sergio A. and {Eatough}, Ralph P. and {Emami}, Razieh and {Falcke}, Heino and {Farah}, Joseph and {Fish}, Vincent L. and {Fomalont}, Ed and {Ford}, H. Alyson and {Fraga-Encinas}, Raquel and {Freeman}, William T. and {Friberg}, Per and {Fromm}, Christian M. and {Fuentes}, Antonio and {Galison}, Peter and {Gammie}, Charles F. and {Garc{\'\i}a}, Roberto and {Gentaz}, Olivier and {Georgiev}, Boris and {Goddi}, Ciriaco and {Gold}, Roman and {G{\'o}mez-Ruiz}, Arturo I. and {G{\'o}mez}, Jos{\'e} L. and {Gu}, Minfeng and {Gurwell}, Mark and {Hada}, Kazuhiro and {Haggard}, Daryl and {Haworth}, Kari and {Hecht}, Michael H. and {Hesper}, Ronald and {Heumann}, Dirk and {Ho}, Luis C. and {Ho}, Paul and {Honma}, Mareki and {Huang}, Chih-Wei L. and {Huang}, Lei and {Hughes}, David H. and {Ikeda}, Shiro and {Impellizzeri}, C.~M. Violette and {Inoue}, Makoto and {Issaoun}, Sara and {James}, David J. and {Jannuzi}, Buell T. and {Janssen}, Michael and {Jeter}, Britton and {Jiang}, Wu and {Jim{\'e}nez-Rosales}, Alejandra and {Johnson}, Michael D. and {Jorstad}, Svetlana and {Joshi}, Abhishek V. and {Jung}, Taehyun and {Karami}, Mansour and {Karuppusamy}, Ramesh and {Kawashima}, Tomohisa and {Keating}, Garrett K. and {Kettenis}, Mark and {Kim}, Dong-Jin and {Kim}, Jae-Young and {Kim}, Jongsoo and {Kim}, Junhan and {Kino}, Motoki and {Koay}, Jun Yi and {Kocherlakota}, Prashant and {Kofuji}, Yutaro and {Koch}, Patrick M. and {Koyama}, Shoko and {Kramer}, Carsten and {Kramer}, Michael and {Krichbaum}, Thomas P. and {Kuo}, Cheng-Yu and {Bella}, Noemi La and {Lauer}, Tod R. and {Lee}, Daeyoung and {Lee}, Sang-Sung and {Leung}, Po Kin and {Levis}, Aviad and {Li}, Zhiyuan and {Lico}, Rocco and {Lindahl}, Greg and {Lindqvist}, Michael and {Lisakov}, Mikhail and {Liu}, Jun and {Liu}, Kuo and {Liuzzo}, Elisabetta and {Lo}, Wen-Ping and {Lobanov}, Andrei P. and {Loinard}, Laurent and {Lonsdale}, Colin J. and {Lu}, Ru-Sen and {Mao}, Jirong and {Marchili}, Nicola and {Markoff}, Sera and {Marrone}, Daniel P. and {Marscher}, Alan P. and {Mart{\'\i}-Vidal}, Iv{\'a}n and {Matsushita}, Satoki and {Matthews}, Lynn D. and {Medeiros}, Lia and {Menten}, Karl M. and {Michalik}, Daniel and {Mizuno}, Izumi and {Mizuno}, Yosuke and {Moran}, James M. and {Moriyama}, Kotaro and {Moscibrodzka}, Monika and {M{\"u}ller}, Cornelia and {Mus}, Alejandro and {Musoke}, Gibwa and {Myserlis}, Ioannis and {Nadolski}, Andrew and {Nagai}, Hiroshi and {Nagar}, Neil M. and {Nakamura}, Masanori and {Narayan}, Ramesh and {Narayanan}, Gopal and {Natarajan}, Iniyan and {Nathanail}, Antonios and {Fuentes}, Santiago Navarro and {Neilsen}, Joey and {Neri}, Roberto and {Ni}, Chunchong and {Noutsos}, Aristeidis and {Nowak}, Michael A. and {Oh}, Junghwan and {Okino}, Hiroki and {Olivares}, H{\'e}ctor and {Ortiz-Le{\'o}n}, Gisela N. and {Oyama}, Tomoaki and {Palumbo}, Daniel C.~M. and {Paraschos}, Georgios Filippos and {Park}, Jongho and {Parsons}, Harriet and {Patel}, Nimesh and {Pen}, Ue-Li and {Pesce}, Dominic W. and {Pi{\'e}tu}, Vincent and {Plambeck}, Richard and {PopStefanija}, Aleksandar and {Porth}, Oliver and {P{\"o}tzl}, Felix M. and {Prather}, Ben and {Preciado-L{\'o}pez}, Jorge A. and {Pu}, Hung-Yi and {Ramakrishnan}, Venkatessh and {Rao}, Ramprasad and {Rawlings}, Mark G. and {Raymond}, Alexander W. and {Rezzolla}, Luciano and {Ricarte}, Angelo and {Ripperda}, Bart and {Roelofs}, Freek and {Rogers}, Alan and {Ros}, Eduardo and {Romero-Ca{\~n}izales}, Cristina and {Roshanineshat}, Arash and {Rottmann}, Helge and {Roy}, Alan L. and {Ruiz}, Ignacio and {Ruszczyk}, Chet and {Rygl}, Kazi L.~J. and {S{\'a}nchez}, Salvador and {S{\'a}nchez-Arg{\"u}elles}, David and {S{\'a}nchez-Portal}, Miguel and {Sasada}, Mahito and {Satapathy}, Kaushik and {Savolainen}, Tuomas and {Schloerb}, F. Peter and {Schonfeld}, Jonathan and {Schuster}, Karl-Friedrich and {Shao}, Lijing and {Shen}, Zhiqiang and {Small}, Des and {Sohn}, Bong Won and {SooHoo}, Jason and {Souccar}, Kamal and {Sun}, He and {Tazaki}, Fumie and {Tetarenko}, Alexandra J. and {Tiede}, Paul and {Tilanus}, Remo P.~J. and {Titus}, Michael and {Torne}, Pablo and {Traianou}, Efthalia and {Trent}, Tyler and {Trippe}, Sascha and {Turk}, Matthew and {van Bemmel}, Ilse and {van Langevelde}, Huib Jan and {van Rossum}, Daniel R. and {Vos}, Jesse and {Wagner}, Jan and {Ward-Thompson}, Derek and {Wardle}, John and {Weintroub}, Jonathan and {Wex}, Norbert and {Wharton}, Robert and {Wielgus}, Maciek and {Wiik}, Kaj and {Witzel}, Gunther and {Wondrak}, Michael F. and {Wong}, George N. and {Wu}, Qingwen and {Yamaguchi}, Paul and {Yoon}, Doosoo and {Young}, Andr{\'e} and {Young}, Ken and {Younsi}, Ziri and {Yuan}, Feng and {Yuan}, Ye-Fei and {Zensus}, J. Anton and {Zhang}, Shuo and {Zhao}, Guang-Yao and {Zhao}, Shan-Shan and {Chang}, Dominic O.},
        title = "{First Sagittarius A* Event Horizon Telescope Results. IV. Variability, Morphology, and Black Hole Mass}",
      journal = {\apjl},
         year = 2022,
        month = may,
       volume = {930},
       number = {2},
          eid = {L15},
        pages = {L15},
          doi = {10.3847/2041-8213/ac6736},
       adsurl = {https://ui.adsabs.harvard.edu/abs/2022ApJ...930L..15A}
}

@BOOK{TMS,
   author = {{Thompson}, A.~R. and {Moran}, J.~M. and {Swenson}, Jr., G.~W.
	},
    title = "{Interferometry and Synthesis in Radio Astronomy, 3rd Edition}",
publisher = {Springer International Publishing},
booktitle = {Interferometry and Synthesis in Radio Astronomy, by A.~Richard Thompson, James M.~Moran, and George W.~Swenson, Jr.~3rd ed.~Springer, 2017.},
     year = 2017,
      doi = {10.1007/978-3-319-44431-4},
   adsurl = {http://adsabs.harvard.edu/abs/2017isra.book.....T}
}

@ARTICLE{Kim_2018,
       author = {{Kim}, J. -Y. and {Krichbaum}, T.~P. and {Lu}, R. -S. and {Ros}, E. and {Bach}, U. and {Bremer}, M. and {de Vicente}, P. and {Lindqvist}, M. and {Zensus}, J.~A.},
        title = "{The limb-brightened jet of M87 down to the 7 Schwarzschild radii scale}",
      journal = {\aap},
         year = 2018,
        month = sep,
       volume = {616},
          eid = {A188},
        pages = {A188},
          doi = {10.1051/0004-6361/201832921},
archivePrefix = {arXiv},
       eprint = {1805.02478},
 primaryClass = {astro-ph.GA},
       adsurl = {https://ui.adsabs.harvard.edu/abs/2018A&A...616A.188K}
}

@ARTICLE{Walker_2018,
       author = {{Walker}, R. Craig and {Hardee}, Philip E. and {Davies}, Frederick B. and {Ly}, Chun and {Junor}, William},
        title = "{The Structure and Dynamics of the Subparsec Jet in M87 Based on 50 VLBA Observations over 17 Years at 43 GHz}",
      journal = {\apj},
         year = 2018,
        month = mar,
       volume = {855},
       number = {2},
          eid = {128},
        pages = {128},
          doi = {10.3847/1538-4357/aaafcc},
archivePrefix = {arXiv},
       eprint = {1802.06166},
 primaryClass = {astro-ph.HE},
       adsurl = {https://ui.adsabs.harvard.edu/abs/2018ApJ...855..128W}
}

@ARTICLE{Levis_tomography,
 journal = {arXiv},
  doi = {10.48550/ARXIV.2204.03715},
  url = {https://arxiv.org/abs/2204.03715},
  author = {Levis, Aviad and Srinivasan, Pratul P. and Chael, Andrew A. and Ng, Ren and Bouman, Katherine L.},
  title = {Gravitationally Lensed Black Hole Emission Tomography},
  publisher = {arXiv},
  year = {2022},
  copyright = {Creative Commons Attribution 4.0 International}
}

@ARTICLE{Goddi_2021,
       author = {{Goddi}, Ciriaco and {Mart{\'\i}-Vidal}, Iv{\'a}n and {Messias}, Hugo and {Bower}, Geoffrey C. and {Broderick}, Avery E. and {Dexter}, Jason and {Marrone}, Daniel P. and {Moscibrodzka}, Monika and {Nagai}, Hiroshi and {Algaba}, Juan Carlos and {Asada}, Keiichi and {Crew}, Geoffrey B. and {G{\'o}mez}, Jos{\'e} L. and {Impellizzeri}, C.~M. Violette and {Janssen}, Michael and {Kadler}, Matthias and {Krichbaum}, Thomas P. and {Lico}, Rocco and {Matthews}, Lynn D. and {Nathanail}, Antonios and {Ricarte}, Angelo and {Ros}, Eduardo and {Younsi}, Ziri and {Akiyama}, Kazunori and {Alberdi}, Antxon and {Alef}, Walter and {Anantua}, Richard and {Azulay}, Rebecca and {Baczko}, Anne-Kathrin and {Ball}, David and {Balokovi{\'c}}, Mislav and {Barrett}, John and {Benson}, Bradford A. and {Bintley}, Dan and {Blackburn}, Lindy and {Blundell}, Raymond and {Boland}, Wilfred and {Bouman}, Katherine L. and {Boyce}, Hope and {Bremer}, Michael and {Brinkerink}, Christiaan D. and {Brissenden}, Roger and {Britzen}, Silke and {Broguiere}, Dominique and {Bronzwaer}, Thomas and {Byun}, Do-Young and {Carlstrom}, John E. and {Chael}, Andrew and {Chan}, Chi-kwan and {Chatterjee}, Shami and {Chatterjee}, Koushik and {Chen}, Ming-Tang and {Chen}, Yongjun and {Chesler}, Paul M. and {Cho}, Ilje and {Christian}, Pierre and {Conway}, John E. and {Cordes}, James M. and {Crawford}, Thomas M. and {Cruz-Osorio}, Alejandro and {Cui}, Yuzhu and {Davelaar}, Jordy and {De Laurentis}, Mariafelicia and {Deane}, Roger and {Dempsey}, Jessica and {Desvignes}, Gregory and {Doeleman}, Sheperd S. and {Eatough}, Ralph P. and {Falcke}, Heino and {Farah}, Joseph and {Fish}, Vincent L. and {Fomalont}, Ed and {Ford}, H. Alyson and {Fraga-Encinas}, Raquel and {Freeman}, William T. and {Friberg}, Per and {Fromm}, Christian M. and {Fuentes}, Antonio and {Galison}, Peter and {Gammie}, Charles F. and {Garc{\'\i}a}, Roberto and {Gentaz}, Olivier and {Georgiev}, Boris and {Gold}, Roman and {G{\'o}mez-Ruiz}, Arturo I. and {Gu}, Minfeng and {Gurwell}, Mark and {Hada}, Kazuhiro and {Haggard}, Daryl and {Hecht}, Michael H. and {Hesper}, Ronald and {Ho}, Luis C. and {Ho}, Paul and {Honma}, Mareki and {Huang}, Chih-Wei L. and {Huang}, Lei and {Hughes}, David H. and {Inoue}, Makoto and {Issaoun}, Sara and {James}, David J. and {Jannuzi}, Buell T. and {Jeter}, Britton and {Jiang}, Wu and {Jimenez-Rosales}, Alejandra and {Johnson}, Michael D. and {Jorstad}, Svetlana and {Jung}, Taehyun and {Karami}, Mansour and {Karuppusamy}, Ramesh and {Kawashima}, Tomohisa and {Keating}, Garrett K. and {Kettenis}, Mark and {Kim}, Dong-Jin and {Kim}, Jae-Young and {Kim}, Jongsoo and {Kim}, Junhan and {Kino}, Motoki and {Koay}, Jun Yi and {Kofuji}, Yutaro and {Koch}, Patrick M. and {Koyama}, Shoko and {Kramer}, Michael and {Kramer}, Carsten and {Kuo}, Cheng-Yu and {Lauer}, Tod R. and {Lee}, Sang-Sung and {Levis}, Aviad and {Li}, Yan-Rong and {Li}, Zhiyuan and {Lindqvist}, Michael and {Lindahl}, Greg and {Liu}, Jun and {Liu}, Kuo and {Liuzzo}, Elisabetta and {Lo}, Wen-Ping and {Lobanov}, Andrei P. and {Loinard}, Laurent and {Lonsdale}, Colin and {Lu}, Ru-Sen and {MacDonald}, Nicholas R. and {Mao}, Jirong and {Marchili}, Nicola and {Markoff}, Sera and {Marscher}, Alan P. and {Matsushita}, Satoki and {Medeiros}, Lia and {Menten}, Karl M. and {Mizuno}, Izumi and {Mizuno}, Yosuke and {Moran}, James M. and {Moriyama}, Kotaro and {M{\"u}ller}, Cornelia and {Musoke}, Gibwa and {Mej{\'\i}as}, Alejandro Mus and {Nagar}, Neil M. and {Nakamura}, Masanori and {Narayan}, Ramesh and {Narayanan}, Gopal and {Natarajan}, Iniyan and {Neilsen}, Joey and {Neri}, Roberto and {Ni}, Chunchong and {Noutsos}, Aristeidis and {Nowak}, Michael A. and {Okino}, Hiroki and {Olivares}, H{\'e}ctor and {Ortiz-Le{\'o}n}, Gisela N. and {Oyama}, Tomoaki and {{\"O}zel}, Feryal and {Palumbo}, Daniel C.~M. and {Park}, Jongho and {Patel}, Nimesh and {Pen}, Ue-Li and {Pesce}, Dominic W. and {Pi{\'e}tu}, Vincent and {Plambeck}, Richard and {PopStefanija}, Aleksandar and {Porth}, Oliver and {P{\"o}tzl}, Felix M. and {Prather}, Ben and {Preciado-L{\'o}pez}, Jorge A. and {Psaltis}, Dimitrios and {Pu}, Hung-Yi and {Ramakrishnan}, Venkatessh and {Rao}, Ramprasad and {Rawlings}, Mark G. and {Raymond}, Alexander W. and {Rezzolla}, Luciano and {Ripperda}, Bart and {Roelofs}, Freek and {Rogers}, Alan and {Rose}, Mel and {Roshanineshat}, Arash and {Rottmann}, Helge and {Roy}, Alan L. and {Ruszczyk}, Chet and {Rygl}, Kazi L.~J. and {S{\'a}nchez}, Salvador and {S{\'a}nchez-Arguelles}, David and {Sasada}, Mahito},
        title = "{Polarimetric Properties of Event Horizon Telescope Targets from ALMA}",
      journal = {\apjl},
         year = 2021,
        month = mar,
       volume = {910},
       number = {1},
          eid = {L14},
        pages = {L14},
          doi = {10.3847/2041-8213/abee6a},
archivePrefix = {arXiv},
       eprint = {2105.02272},
 primaryClass = {astro-ph.GA},
       adsurl = {https://ui.adsabs.harvard.edu/abs/2021ApJ...910L..14G}
}

@article{Gelles_2021_published,
  title = {Polarized image of equatorial emission in the Kerr geometry},
  author = {Gelles, Zachary and Himwich, Elizabeth and Johnson, Michael D. and Palumbo, Daniel C. M.},
  journal = {Phys. Rev. D},
  volume = {104},
  issue = {4},
  pages = {044060},
  numpages = {25},
  year = {2021},
  month = {Aug},
  publisher = {American Physical Society},
  doi = {10.1103/PhysRevD.104.044060},
  url = {https://link.aps.org/doi/10.1103/PhysRevD.104.044060}
}

@ARTICLE{Chang_Krang_2024,
       author = {{Chang}, Dominic},
        title = "{Krang: Kerr Raytracer for Analytic Null Geodesics}",
      journal = {The Journal of Open Source Software},
         year = 2024,
        month = oct,
       volume = {9},
       number = {102},
          eid = {7273},
        pages = {7273},
          doi = {10.21105/joss.07273},
       adsurl = {https://ui.adsabs.harvard.edu/abs/2024JOSS....9.7273C}
}

@ARTICLE{Mosci_2016,
   author = {{Mo{\'s}cibrodzka}, M. and {Falcke}, H. and {Shiokawa}, H.},
    title = "{General relativistic magnetohydrodynamical simulations of the jet in M 87}",
  journal = {\aap},
archivePrefix = "arXiv",
   eprint = {1510.07243},
 primaryClass = "astro-ph.HE",
     year = 2016,
    month = feb,
   volume = 586,
      eid = {A38},
    pages = {A38},
      doi = {10.1051/0004-6361/201526630},
   adsurl = {http://adsabs.harvard.edu/abs/2016A%26A...586A..38M}
}

@ARTICLE{Palumbo_2025,
       author = {{Palumbo}, Daniel C.~M.},
        title = "{Supermassive Black Hole Spin Constraints from Polarimetry in an Equatorial Disk Model}",
      journal = {\apjl},
         year = 2025,
        month = jan,
       volume = {978},
       number = {1},
          eid = {L4},
        pages = {L4},
          doi = {10.3847/2041-8213/ad9bb4},
archivePrefix = {arXiv},
       eprint = {2412.06719},
 primaryClass = {astro-ph.HE},
       adsurl = {https://ui.adsabs.harvard.edu/abs/2025ApJ...978L...4P}
}

@ARTICLE{Hou_2025,
       author = {{Hou}, Yehui and {Huang}, Jiewei and {Guo}, Minyong and {Mizuno}, Yosuke and {Chen}, Bin},
        title = "{Near-horizon Polarization as a Diagnostic of Black Hole Spacetime}",
      journal = {\apjl},
         year = 2025,
        month = aug,
       volume = {988},
       number = {2},
          eid = {L51},
        pages = {L51},
          doi = {10.3847/2041-8213/adee09},
archivePrefix = {arXiv},
       eprint = {2409.07248},
 primaryClass = {gr-qc},
       adsurl = {https://ui.adsabs.harvard.edu/abs/2025ApJ...988L..51H}
}

@ARTICLE{Wong_2025_published,
       author = {{Wong}, George N. and {Chael}, Andrew and {Lupsasca}, Alexandru and {Quataert}, Eliot},
        title = "{Black Hole Polarimetry. II. The Connection between Spin and Polarization}",
      journal = {\apj},
         year = 2026,
        month = jan,
       volume = {997},
       number = {1},
          eid = {113},
        pages = {113},
          doi = {10.3847/1538-4357/ae20f3},
archivePrefix = {arXiv},
       eprint = {2509.22639},
 primaryClass = {astro-ph.HE},
       adsurl = {https://ui.adsabs.harvard.edu/abs/2026ApJ...997..113W}
}

@ARTICLE{BZ_1977,
       author = {{Blandford}, R.~D. and {Znajek}, R.~L.},
        title = "{Electromagnetic extraction of energy from Kerr black holes.}",
      journal = {\mnras},
         year = 1977,
        month = may,
       volume = {179},
        pages = {433-456},
          doi = {10.1093/mnras/179.3.433},
       adsurl = {https://ui.adsabs.harvard.edu/abs/1977MNRAS.179..433B}
}

@ARTICLE{Gralla_2019_photonrings,
       author = {{Gralla}, Samuel E. and {Holz}, Daniel E. and {Wald}, Robert M.},
        title = "{Black hole shadows, photon rings, and lensing rings}",
      journal = {\prd},
         year = 2019,
        month = jul,
       volume = {100},
       number = {2},
          eid = {024018},
        pages = {024018},
          doi = {10.1103/PhysRevD.100.024018},
archivePrefix = {arXiv},
       eprint = {1906.00873},
 primaryClass = {astro-ph.HE},
       adsurl = {https://ui.adsabs.harvard.edu/abs/2019PhRvD.100b4018G}
}

@ARTICLE{Yuan_and_Narayan_2014,
       author = {{Yuan}, Feng and {Narayan}, Ramesh},
        title = "{Hot Accretion Flows Around Black Holes}",
      journal = {\araa},
         year = 2014,
        month = aug,
       volume = {52},
        pages = {529-588},
          doi = {10.1146/annurev-astro-082812-141003},
archivePrefix = {arXiv},
       eprint = {1401.0586},
 primaryClass = {astro-ph.HE},
       adsurl = {https://ui.adsabs.harvard.edu/abs/2014ARA&A..52..529Y}
}
\end{document}